\documentclass[pre,showpacs]{revtex4}
\usepackage{latexsym}
\usepackage{graphicx}

\newcommand{\ba}{\begin{eqnarray}}
\newcommand{\be}{\begin{equation}}
\newcommand{\ea}{\end{eqnarray}}
\newcommand{\ee}{\end{equation}}
\renewcommand{\k}{{\bf k}}
\newcommand{\kap}{{\mbox{\boldmath$\kappa$}}}
\newcommand{\Nabla}{{\mbox{\boldmath$\nabla$}}}
\newcommand{\q}{{\bf q}}
\renewcommand{\r}{{\bf r}}
\newcommand{\Ev}{E_{\mathrm{vac}}}

\newcommand{\iv}{{\mathrm{in}}}
\newcommand{\ou}{{\mathrm{out}}}

\newcommand{\ignore}[1]{{}}

\begin{document}

\title{Response theory for time-resolved second-harmonic generation\\
and two-photon photoemission}

\author{C. Timm}
\email{timm@physik.fu-berlin.de.}
\author{K. H. Bennemann}
\affiliation{Institut f\"ur Theoretische Physik, Freie Universit\"at
Berlin, Arnimallee 14, D-14195 Berlin, Germany}

\date{August 19, 2003}

\begin{abstract}
A unified response theory for the time-resolved nonlinear light generation
and two-photon photoemission (2PPE) from metal surfaces is presented. The
theory allows to describe the dependence of the nonlinear optical response
and the photoelectron yield, respectively, on the time dependence of the
exciting light field. Quantum-mechanical interference effects affect
the results
significantly. Contributions to 2PPE due to the optical nonlinearity of the
surface region are derived and shown to be relevant close to a plasmon
resonance. The interplay between pulse shape, relaxation times of excited
electrons, and band structure is analyzed directly in the time domain.
While our theory works for arbitrary pulse shapes, we mainly focus on the
case of two pulses of the same mean frequency. Difficulties in extracting
relaxation rates from pump-probe experiments are discussed, for example due
to the effect of detuning of intermediate states on the interference. The
theory also allows to determine the range of validity of the optical Bloch
equations and of semiclassical rate equations, respectively. Finally, we
discuss how collective plasma excitations affect the nonlinear
optical response and 2PPE.
\end{abstract}


\maketitle

\section{Introduction}
\label{sec.intro}

During the last decade time-resolved spectroscopy of condensed-matter
systems has become a very active area of experimental research
\cite{Ste92,HMKB97,Sim98,KSR98,Gud99,Schoe,Aesch1,Aesch2,Wolf1,%
Hoef97,Cao97,Petek,Wolf2,Leh99,Kno00,GWH00,SBW02,ONP02,Gerb,Ae.cluster,%
Petekrev,Fausterrev}. This is mainly due to the progress in experimental
technique, in particular the ability to create ultra-short laser pulses
with a duration of the order of a few femtoseconds \cite{femto}. Since this
is similar to the relaxation times of excited electrons and collective
excitations in solids, these experiments allow to study non-equilibrium
physics, {\it e.g.}, the time evolution of excited electrons before and
during thermalization. Of particular interest are non-linear techniques
such as time-resolved sum-frequency generation (SFG) and two-photon
photoemission (2PPE), which are sensitive to excited electron states
\cite{Shen}. A theoretical understanding of these processes is crucial.
Petek and Ogawa \cite{Petekrev} noted in 1997 that a theory for
time-resolved 2PPE is still lacking, and, despite the efforts of many
theorists, much remains to be done. The situation for SFG is similar. The
construction of such a theory is a formidable task---the main problems are
(a) the desription of the time-dependent response and (b) the treatment of
the surface. Our main concern is with the first point. A simplified
description of the surface using Fresnel factors has been employed
successfully to describe SFG from metals \cite{SMD87,HBB94,Luce97,LuceLDA}.
A detailed discussion of boundary conditions at the surface, focusing on
the nonlinear optical response of magnetic systems, can be found in
Ref.~\cite{AtK02}.

In the present paper we discuss the electronic processes taking place
during time-resolved SFG (in particular second-harmonic generation, SHG)
and 2PPE and derive the dependence of the SFG light intensity and the 2PPE
photoelectron yield on the time dependence of the exciting laser field. We
show that most effects observed for time-resolved 2PPE appear similarly for
SFG, such as their dependence on energy relaxation, dephasing, and detuning
of intermediate states. Other examples are the enhancement of the response
due to collective excitations and the sensitivity regarding the ultra-fast
spin-dependent relaxation. We develop a unified time-dependent response
theory for SFG and 2PPE, starting from the self-consistent field approach
of Ehrenreich and Cohen \cite{EC,HB}, which can be applied to specific
materials described by their band structure, relaxation rates, and dipole
matrix elements. For illustration, we apply the theory to a generic
tight-binding model for a metal to study interference effects in both
pump-probe single-color SFG and 2PPE and their dependence on relaxation
rates and detuning. We exhibit the strong similarities between both methods.


\begin{figure}[ht]
\includegraphics[width=5.00cm]{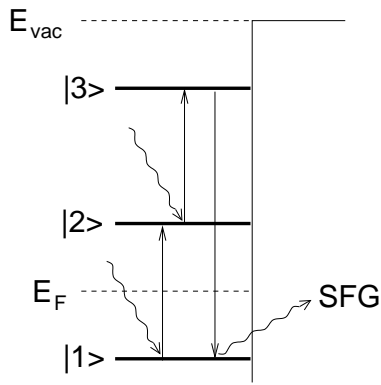}
\caption{Simplified representation of sum-frequency generation (SFG). 
$E_F$ is the Fermi energy and $\Ev$ is the vacuum energy.
In SFG two photons of frequencies $\omega_1$ and $\omega_2$ are absorbed
by electrons in states $|1\rangle$ and $|2\rangle$
and a single photon of frequency $\omega_1+\omega_2$ is emitted due to the
electronic transition $|3\rangle\to|1\rangle$.
The two photons may be provided by one or
two laser pulses. Note, whether the two photons are predominantly
absorbed at nearly the same time or with some delay depends on the
shape and width of the pulse(s). A more careful analysis of the absorption
process on the basis
of response theory is given in Sec.~\protect\ref{ss.21}.}
\label{fig1}
\end{figure}

In SFG \cite{Ste92,HMKB97,Sim98,KSR98,Gud99} electrons are excited by
absorbing two photons and they subsequently emit a single photon at the sum
frequency. In Fig.~\ref{fig1} we illustrate the type of process yielding
SFG. For simplicity we talk about SFG in the following, although
\emph{difference}-frequency generation is automatically included in our
theory. Time-resolved measurements \cite{Ste92,HMKB97,Sim98,KSR98,Gud99}
usually employ the pump-probe technique, where two laser pulses of the same
(single-color) or different (two-color) frequency are applied with a time
delay $\Delta T$ between them. This time delay controls the time between
the two absorptions and thus the relaxation dynamics of the electron in the
intermediate state $|2\rangle$ is crucial, see Fig.~\ref{fig1}. SFG is
strongly surface-sensitive, since the SFG response of the bulk of an
inversion-symmetric crystal vanishes in the dipole approximation. The
inversion symmetry can also be broken by nanostructures. The most
important case of SFG is second-harmonic generation (SHG), where the
electrons are excited by approximately monochromatic light of frequency
$\omega$ and light of frequency $2\omega$ is detected. Note, in the case of
ultra-short laser pulses the spectrum is necessarily broadened and a full
treatment of SFG is required even for these single-color experiments. Also
note that a single laser pulse, depending on its duration and shape,
involves time-delayed absorptions.

\begin{figure}[ht]
\includegraphics[width=6.00cm]{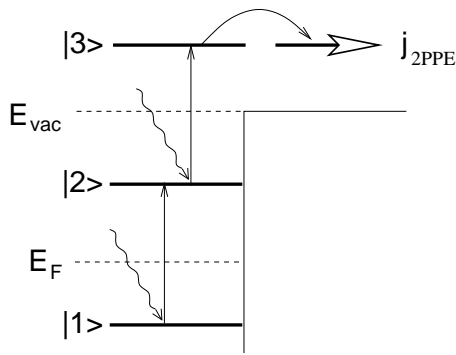}
\caption{Simplified representation of two-photon photoemission (2PPE),
where $E_F$ is the Fermi energy and $\Ev$ the vacuum energy.
Here, two photons of frequencies $\omega_1$ and $\omega_2$ (out of the same
or different pulses) are absorbed by electrons in states $|1\rangle$ and
$|2\rangle$, respectively.
Compared to SFG, Fig.~\protect\ref{fig1}(a), the excitation energy is now
so large that electrons are excited above $\Ev$
and can leave the solid. The open arrow denotes electrons leaving the
crystal. Note, SFG is also possible due to a transition
$|3\rangle\to|1\rangle$. However, the SFG intensity may be small, since it
involves more dipole matrix elements, as we discuss below.}
\label{fig2}
\end{figure}

Time-resolved 2PPE experiments of metal surfaces
\cite{Schoe,Aesch1,Aesch2,Wolf1,Hoef97,Cao97,Petek,Wolf2,Leh99,Kno00,%
GWH00,SBW02,ONP02} as well as of clusters \cite{Gerb,Ae.cluster} employing
the pump-probe technique have been performed more often than time-resolved
SFG. Reviews can be found in Refs.~\cite{Petekrev} and \cite{Fausterrev}.
Figure \ref{fig2} shows a sketch of the processes yielding 2PPE. An
electron is excited above the vacuum energy $\Ev$ due to the absorption of
two photons. The interplay between the relaxation of the electrons in
intermediate states and the time between the two absorptions will determine
the resulting photoelectron current. The probability of electrons above the
vacuum level actually leaving the solid is also crucial. The limited mean
free path of the electrons makes photoemission surface-sensitive, but in
general less than in the case of SFG. In both SFG and 2PPE interference
effects \cite{Ste92,Sim98,Aesch2,Cao97,Petek,Wolf2} appear, which our
theory allows to study. Of course, these interference effects are expected
to depend on the pulse shapes.

The response theory presented here goes beyond previous theoretical
treatments of ultra-fast processes \cite{rem.slow} in SFG and 2PPE in
metals, which mainly fall into four classes: (a) density functional theory
and approaches based thereon
\cite{Lie89,Ull97,LuceLDA,Lie99,Koh99,WDS,Camp}, (b) rate equations
\cite{Kno00,Rol1,Rol2}, (c) optical Bloch equations
\cite{Wolf1,Petek,Loudon}, and (d) perturbative methods
\cite{HB,HBB94,PPK00,Sha00}. At first, density functional theory has been
applied in the time-dependent local-density approximation for jellium
models \cite{Lie89,Ull97,Lie99,Koh99}. In the jellium approximation one
ignores the potential of the ion cores and, consequently, any
band-structure effects. Thus this approach is not suitable if single bands
or surface states or quantum-well states in thin films are important. On
the other hand, collective excitations are usually described rather well
\cite{Lieplas}. Going beyond the jellium model, Luce and Bennemann have
employed the local density approximation to calculate dipole matrix
elements as they enter also in our approach \cite{LuceLDA}. Additionally
taking excited states into account within the {\it GW\/} approximation,
Sch\"one {\it et al.} \cite{WDS} have calculated electronic lifetimes. Hole
dynamics have also been studied with density functional methods
\cite{Camp}.

However, one would like to gain more general physical insight than the
numerical results can provide. To this end one may consider rate equations
for the occupation of excited states, {\it e.g.}, the Boltzmann equation
\cite{Kno00,Rol1,Rol2}. This approach allows to incorporate important
effects such as secondary electrons due to relaxation from higher-energy
states and to Auger processes as well as transport into the bulk
\cite{Kno00,Rol1,Rol2}. However, rate equations neglect the electric
polarization of the electron gas, its dephasing, and any quantum-mechanical
interference effects, resulting from the superposition of the laser field
and the induced fields. To include these effects one has to solve the
equation of motion for the entire density matrix $\rho$, not only for its
diagonal components, \textit{i.e.}, the occupations. This can be done in
response theory. Its simplest form yields the optical Bloch equations: The
system is modelled by a small number of levels and the von Neumann equation
of motion (master equation) for the density matrix is integrated
numerically \cite{Loudon,Wolf1,Petek}. However, this approach is
limited to a small number of levels so that a realistic band
structure cannot be described. Furthermore, many-particle effects like
collective excitations are not included.

On the other hand, the response theory presented here does include the band
structure and collective excitations. It generalizes the theory of H\"ubner
and Bennemann \cite{HB} to SFG due to incident light of arbitrary time
dependence and spectrum. The previous theory \cite{HB} has been used
successfully for SHG from metal surfaces, thin films, quantum wells, and
metallic monolayers due to {\it continuous-wave}, monochromatic light
\cite{HB,HBB94,Luce97,LuceLDA,NOLI,Luce96,Luce98,AnH02}. However, the
dependence of SHG on the pulse shape and the effect of energy relaxation
and dephasing were not discussed. We also derive the response expressions
for time-dependent 2PPE within the same framework. Since our theory is
explicitly formulated for continuous bands, it can also serve as a basis
for the discussion of the averaging effects due to bands of finite width
discussed in a more heuristic framework using optical Bloch equations for
discrete levels in Ref.~\cite{Weida}. Since the full time or frequency
dependence is included, effects of frequency broadening of short pulses and
of finite frequency resolution of the detector (for SFG) \cite{Weida} are
easily studied.

Our theory employs a generalized self-consistent-field approach
\cite{EC,HB}, which is equivalent to the random-phase approximation (RPA)
\cite{Dasgupta,Lind54,HG70,Ch98,Mahan}.
We employ the electric-dipole approximation, which is valid for small wave
vector $\q$ of the electromagnetic field and has been used successfully to
describe SHG from metal surfaces
\cite{HBB94,Luce97,LuceLDA,NOLI,Luce96,Luce98,AnH02,Shala}. This is
reasonable, since the skin depth, which is the length scale of field
changes, is about one order of magnitude larger than the lattice constant.
One has to take care in interpreting SFG experiments for
inversion-symmetric crystals, since the surface contribution only dominates
over higher multipole bulk contributions for surfaces of low symmetry
\cite{HBB94,HB}. Similar in spirit to our response theory, Ueba \cite{Ueba}
has studied continuous-wave 2PPE from metal surfaces, Pedersen \textit{et
al}.\ \cite{PPK00} have considered continous-wave SHG from metallic quantum
wells, and Shahbazyan and Perakis \cite{Sha00} have developed a
time-dependent, but {\it linear\/} response theory for metallic
nanoparticles.

It is important to understand that at the surface of a metal, in thin
films, and in nanostructures the light couples to collective plasma
excitations. The field within the metal is of course not purely transverse
\cite{Barton,Board}. Its transverse and longitudinal components couple with
the conduction electrons to form plasmon-polaritons and plasmons,
respectively \cite{Board}. The (longitudinal) plasmon modes only decouple
from the applied field for a structureless jellium model of the solid
\cite{Barton,Board}. However, we consider a more realistic model that
incorporates the crystal structure. Also, we will see that the induced
nonlinear polarization couples to (longitudinal) plasmon modes.

On general grounds one may expect that the discussion of the intimate
relationship between 2PPE and SFG also helps to understand the dependence of
2PPE on light polarization. It has been shown that the light-polarization
dependence of SFG is important for the analysis of the electronic structure
and magnetism \cite{Oxf}.


The organization of the remainder of this paper is as follows: We first
summarize the response theory for SFG and 2PPE in Sec.~\ref{sec.theory}.
This lays the ground for our discussion in Sec.~\ref{sec.disc} of
time-dependent SFG and 2PPE. Details of the response theory are given in
appendices \ref{app.a} and \ref{app.b}.

\section{Response theory}
\label{sec.theory}

\subsection{Sum-frequency generation}
\label{ss.21}

We first outline the response theory for SFG. We consider a semi-infinite
solid with single-particle states $|\k_\| l\rangle$ with energies $E_{\k_\|
l}$ described by the momentum $\k_\|$ parallel to the surface, which is
assumed to be perpendicular to the $z$ direction, and a set of additional
quantum numbers $l$. For bulk states, which may be affected by the surface
but are not localized close to it, the composite band index is
$l=(k_z,\nu,\sigma)$, where $k_z$ is the $z$ momentum component \emph{in
the bulk}, $\nu$ is a band index, and $\sigma$ is the spin quantum number.
$k_z$ has a continuous spectrum. On the other hand, for states localized at
the surface, $l$ is discrete. Examples are image-potential states,
adsorbate states, quantum-well states in a thin overlayer, and proper
surface states.

Part of the electron-electron interaction is included by the
self-consistent-field approximation or RPA \cite{EC,Dasgupta}. The remaining
electron-electron scattering is approximately taken into account by
inserting phenomenological relaxation rates \cite{PS00} into the
single-electron Green functions and by shifting the band energies
$E_{\k_\| l}$ \cite{Louisell}. We assume that $E_{\k_\| l}$
are quasiparticle energies containing these shifts. Note, the electron-phonon
interaction only becomes relevant on longer time scales and is not
considered here \cite{rem.slow}. Also, \emph{intra}band contributions to the
response are not considered for simplicity, which is reasonable at optical
frequencies.

The electrons are coupled to the effective electric field ${\bf E}$ within
the solid through a dipolar interaction term (for simplicity we assume that
the dipole coupling dominates). The optically induced polarization ${\bf
P}$ within the solid is expanded in orders of the electric field ${\bf E}$.
The {\it linear\/} response is given by
\be
P_i^{(1)}(\q,t) = \frac{1}{2\pi} \sum_{q'_z}
  \int dt'\:\chi_{ij}(\q,\q';t-t')\,E_j(\q',t') ,
\label{2.P1.2}
\ee
where $\chi_{ij}$ is the linear susceptibility, $\q=(\q_\|,q_z)$, and
$\q_\|=\q'_\|$ due to conservation of momentum parallel to the surface.
Summation over repeated indices is always implied. The non-conservation of
$q_z$ is explicitly taken into account.

\begin{figure}[ht]
\includegraphics[width=4.00cm]{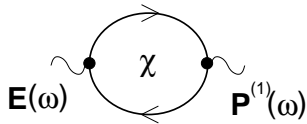}
\caption{The Feynman diagram for the linear susceptibility $\chi$ relating
the linear polarization ${\bf P}^{(1)}$ to the effective electric field
${\bf E}$, \textit{cf}.\ Eq.~(\protect\ref{2.P1.2}).
The solid lines in the diagrams are to be understood as
electronic Matsubara-Green functions containing relaxation rates
$\Gamma$. The dots ($\bullet$) denote dipole matrix elements $\mathbf{D}$.}
\label{fig.chi1}
\end{figure}

We assume throughout that the photon momentum $\q$ is small compared to the
dimensions of the Brillouin zone and that the band energies, relaxation
rates, and transition matrix elements change slowly with momentum so that
the difference between the parallel crystal momentum of an electron before
and after the interaction, $\k_\|$ and $\k_\|'$, respectively, can be ignored.
If we further neglect the frequency dependence of the transition matrix
elements the self-consistent-field approach gives the time-dependent linear
susceptibility
\ba
\lefteqn{
\chi_{ij}(\q,\q';t-t')
  = \frac{e^2}{v} \frac{2\pi i}{\hbar}\,\Theta(t-t')
  } \nonumber \\
& & \quad{}\times \sum_{\k_\|} \sum_{ll'}
  D^i_{\k_\| l';\k_\| l}(-q_z)\,
  D^j_{\k_\| l;\k_\| l'}(q_z') \:
  \nonumber \\
& & \quad{}\times [f(E_{\k_\| l'})-f(E_{\k_\| l})]\,
  \exp\!\left[i \frac{E_{\k_\| l'}-E_{\k_\| l}}
    {\hbar}(t-t')\right]
  \nonumber \\
& & \quad{}\times
  \exp[-\Gamma_{\k_\| l;\k_\| l'}(t-t')] ,\!
\label{2.chi1.3}
\ea
where ${v}$ is the volume of the system. Note, the last two factors
explicitly describe the oscillations and decay of the linear induced
polarization. In the dipole approximation the
transition matrix elements are
\be
{\bf D}_{\k_\| l;\k_\| l'}(q_z) \equiv
  \langle \k_\| l|\, {\bf r}\, |\k_\| l'\rangle .
\ee
The matrix elements are given without approximations in App.~\ref{app.a}.
The linear susceptibility is represented
by the usual Feynman diagram shown in Fig.~\ref{fig.chi1}.

The finite lifetime of electrons due to their interaction enters
Eq.~(\ref{2.chi1.3}) through the {\it dephasing\/} rates $\Gamma_{\k_\|
l;\k'_\| l'}$, which describe the decay of the superposition of states
$|\k_\| l\rangle$ and $|\k'_\| l'\rangle$ and thus of the polarization.
The change of the occupation of states is
described by the {\it energy relaxation\/}
rates $\Gamma_{\k_\| l;\k_\| l} \equiv \tau_{\k_\| l}^{-1}$,
where $\tau_{\k_\| l}$ are the lifetimes. $\Gamma_{\k_\| l;\k_\| l}$ is the
rate of spontaneous transitions out of the state $|\k_\| l\rangle$.
Since the depopulation of the states $|\k_\| l\rangle$ or
$|\k'_\| l'\rangle$ certainly leads to the destruction of the polarization,
the dephasing rates can be expressed in terms of the lifetimes
as \cite{Louisell}
\be
\Gamma_{\k_\| l;\k'_\| l'}
  = \frac{\tau_{\k_\| l}^{-1} + \tau_{\k'_\| l'}^{-1}}{2}
    + \Gamma^{\mathrm{ph}}_{\k_\| l;\k'_\| l'} ,
\label{2.rates1}
\ee
where $\Gamma^{\mathrm{ph}}$ describes additional dephasing.

The induced second-order polarization is given by
\be
P_i^{(2)}(t) = \frac{1}{(2\pi)^2} \int dt_1\,dt_2\:
  \chi^{(2)}_{ijk}(t-t_1,t_1-t_2)\,E_j(t_1)\,E_k(t_2) .
\label{21.P2}
\ee
The second-order susceptibility $\chi^{(2)}$ depends only on two time
differences due to homogeneity in time. Obviously, $|t_1-t_2|$ is the time
interval between the two absorptions. For a single laser pulse this
interval is controlled by the pulse width. For two pulses we expect a
contribution for $|t_1-t_2|$ of the order of the pump-probe delay time
$\Delta T$. Note, the light polarization is characterized by the components
$E_j$.

To express the electric field ${\bf E}$ within the solid in terms of the
applied external light field ${\bf E}_{\mathrm{las}}$ and similarly the
electric field ${\bf E}_{\mathrm{out}}$ of the outgoing light in terms of
the polarization ${\bf P}$ one should employ Fresnel formulae, which are
also of importance for the coupling of the light to collective excitations,
as we discuss below. We do not present the Fresnel formulae here, since
they can be found in the literature \cite{SMD87,HBB94}. See also
Refs.~\cite{Born75,Board} for effective Fresnel factors for systems of
several layers, such as the important case of a coupling prism separated
from the metal by a thin layer of air or vacuum \cite{Otto}. Of course, it
would be of interest to repeat Fresnel's analysis for SFG, in particular
deducing phase shifts etc.

Equation (\ref{21.P2}) is the basis for time-dependent SFG. Clearly the
pulse shape of the applied light described by $\mathbf{E}(t)$ affects the
induced polarization $\mathbf{P}^{(2)}(t)$. Note, for simple pulse shapes
(Gaussian, Lorentzian, rectangular) it is possible to evaluate the
integrals in Eq.~(\ref{21.P2}) further. The light polarization dependence
is controlled by the symmetries of the tensor $\chi^{(2)}_{ijk}$. The
symmetries of $\chi^{(2)}$ for magnetic and nonmagnetic crystals under
monochromatic light have been discussed in Ref.~\cite{Pan}. They are
determined by the symmetry operations that leave the particular surface
invariant. These symmetry arguments are unchanged for general time
dependence of the applied laser field.

The intensity of SFG light is
$I^{(2)}(t) \propto [E^{(2)}_{\mathrm{out}}(t)]^2 \propto [P^{(2)}(t)]^2$.
So far, typical experiments do not resolve the time dependence of the
intensity, but measure the time-integrated SFG yield
\be
{\cal I}^{(2)} \equiv \int dt\: I^{(2)}(t) 
  \propto \int dt\: \left[E^{(2)}_{\mathrm{out}}(t)\right]^2
  \propto \int dt\: \left[P^{(2)}(t)\right]^2  .
\ee
For simplicity we here sum over polarization directions.

Time-resolved SFG may be performed by measuring ${\cal I}^{(2)}(\Delta T)$
as a function of the time delay $\Delta T$ between the applied field
pulses. Omitting surface effects (Fresnel factors) to emphasize the
structure of the results, the yield can be written as
\ba
\lefteqn{
{\cal I}^{(2)}(\Delta T) \propto \int dt\,dt_1\,dt_2\,dt_3\,dt_4\;
  \chi^{(2)}_{ijk}(t-t_1,t_1-t_2) }
  \nonumber \\
& & \quad{}\times 
  \chi^{(2)}_{ilm}(t-t_3,t_3-t_4)\,
  E_j(t_1)\,E_k(t_2)\,E_l(t_3)\,E_m(t_4) ,
\label{21.S2.1}
\ea
which is of fourth order in the incoming light field and thus of second
order in its intensity. As mentioned above, the typical time differences
dominating the response are controlled by the delay $\Delta T$, besides
the pulse durations.

It is useful to write
the second-order polarization $\mathbf{P}^{(2)}$ also in
frequency space,
\be
P_i^{(2)}(\omega) = \int d\omega'\,
  \chi^{(2)}_{ijk}(\omega,\omega')\,E_j(\omega')\,E_k(\omega-\omega') ,
\label{21.P2om}
\ee
where $\chi^{(2)}_{ijk}(t-t_1,t_1-t_2)
= \int d\omega\,d\omega'\, e^{-i\omega (t-t_2)}\, e^{-i\omega' (t_2-t_1)}\,
\chi^{(2)}_{ijk}(\omega,\omega')$ or
\be
  \chi^{(2)}_{ijk}(\omega,\omega') = \frac{1}{4\pi^2}
  \int dt\,dt'\: e^{i\omega t}\, e^{i\omega' t'}\,
  \chi^{(2)}_{ijk}(t+t',-t') .
\label{2.chi21}
\ee
Note that we employ the convention of Eq.~(8) in Ref.~\cite{HB} for the
Fourier transformation. The frequency representation is better suited to
discuss transition \emph{energies}. ${\bf P}^{(2)}$ has components at
the sum of two frequencies of the incoming light. Since the Fourier
transform of the \emph{real} electric field contains positive and negative
frequencies, the difference frequency also appears.

If at the first step we ignore screening effects, then Eqs.~(\ref{21.P2}) and
(\ref{21.S2.1}) only contain the second-order \emph{irreducible}
susceptibility
\ba
\lefteqn{
\chi^{(2)}_{\mathrm{irr};ijk}(\q,\q_1,\q_2;t-t_1,t_1-t_2)
  = -\frac{e^3}{v} \left(\frac{2\pi i}{\hbar}\right)^{\!2}
  \Theta(t-t_1)\,\Theta(t_1-t_2)
  \sum_{\k_\|} \sum_{ll'l''}\,
  D^i_{\k_\| l;\k_\| l''}(-q_z)\,
  \Bigg( D^j_{\k_\| l'';\k_\| l'}(q_{1z}) }
  \nonumber \\
& & \qquad{}\times
  D^k_{\k_\| l';\k_\| l}(q_{2z})\,
  \left[f(E_{\k_\| l})-f(E_{\k_\| l'})\right]\,
  \exp\!\left[i\frac{E_{\k_\| l}-E_{\k_\| l'}}{\hbar}\,(t_1-t_2)\right]\,
  \exp[-\Gamma_{\k_\| l';\k_\| l}(t_1-t_2)]
  \nonumber \\
& & \quad{}- D^j_{\k_\| l';\k_\| l}(q_{1z})
  \, D^k_{\k_\| l'';\k_\| l'}(q_{2z}) \,
  \left[f(E_{\k_\| l'})-f(E_{\k_\| l''})\right]\,
  \exp\!\left[i\frac{E_{\k_\| l'}-E_{\k_\| l''}}{\hbar}\,(t_1-t_2)\right]
  \nonumber \\
& & \qquad{} \times
  \exp[-\Gamma_{\k_\| l'';\k_\| l'}(t_1-t_2)] \Bigg)
  \exp\!\left[i\frac{E_{\k_\| l}-E_{\k_\| l''}}{\hbar}\,(t-t_1)\right]\,
  \exp[-\Gamma_{\k_\| l'';\k_\| l}(t-t_1)] ,
\label{2.chi210}
\ea
which is derived in App.~\ref{app.a}. We neglect the photon momenta
relative to the crystal momentum.
This expression, which forms the basis of our discussion of SFG,
goes beyond the one given in Ref.~\cite{HB} in that it is valid for a
time-dependent and spatially varying laser field.
Furthermore, it includes the transverse response explicitly. Equation
(\ref{2.chi210}) already exhibits the interplay between the time interval
$|t_1-t_2|$ between absorptions, the photon frequencies,
the dephasing times, and the transition frequencies.

\begin{figure}[ht]
\includegraphics[width=6.00cm]{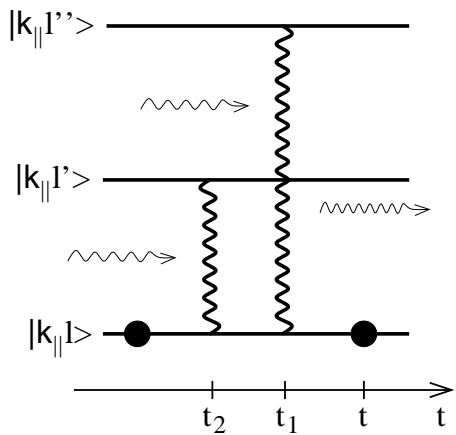}
\caption{Detailed quantum-mechanical interpretation of a process
contributing to SFG.
An electron is excited from a pure state $|\mathbf{k}_\|l\rangle$
in the Fermi sea to a superposition of states $|\mathbf{k}_\|l\rangle$ and
$|\mathbf{k}_\|l''\rangle$ by the absorption of two photons. After the
absorptions the excited electron returns to the original pure state by
emission of a SFG photon at the sum frequency.
The times of the absorptions and the emission are
indicated. The heavy wavy lines denote superpositions resulting from the
absorption of photons (indicated by thin wavy lines with arrows) at $t_2$
and $t_1$, while the black dots represent electrons in pure eigenstates.}
\label{fig.chi2inp}
\end{figure}

We now discuss the physics contained in Eq.~(\ref{2.chi210}) with the help
of Fig.~\ref{fig.chi2inp}. We consider the first of the two terms in
Eq.~(\ref{2.chi210}). The interpretation of the second term is similar
\cite{rem.term2}. The step functions incorporate the time ordering
$t_2<t_1<t$ and thus guarantee causality. The system is in equilibrium
until the first absorption at time $t_2$ creates a {\it superposition\/} of
the two states $|\k_\|l\rangle$ and $|\k_\|l'\rangle$, denoted by the wavy
line in Fig.~\ref{fig.chi2inp}. This important physics is lost in the
interpretation illustrated by Fig.~\ref{fig1}. The Fermi functions make
sure that one of the states is initially occupied and the other is empty.
Let us say state $|\k_\|l\rangle$ is occupied. Since the system is in a
superposition of two eigenstates, the polarization oscillates with the
frequency $(E_{\k_\| l}-E_{\k_\| l'})/\hbar$, as follows from the first
exponential in the parentheses in Eq.~(\ref{2.chi210}). Such superpositions
are described by the {\it off-diagonal\/} components of the density matrix
\cite{rem.linear}. The {\it diagonal\/} components denoting the occupation
numbers of states are \emph{not} changed by a single absorption. The
superposition decays with the dephasing rate $\Gamma_{\k_\| l';\k_\| l}$
associated with this transition, making it clear why the dephasing rates
rather than the energy relaxation rates dominate the response. A second
absorption at the later time \cite{rem.simul} $t_1$ changes the state into
a superposition of the originally occupied state and the state
$|\k_\|l''\rangle$ with its own characteristic oscillation frequency
$(E_{\k_\| l}-E_{\k_\| l''})/\hbar$ and dephasing rate. This oscillating
polarization can emit a photon at that frequency. After the emission the
electron is again in the pure eigenstate $|\k_\|l\rangle$. The nonlinear
susceptibility in Eq.~(\ref{2.chi210}) contains a sum over many
contributions of this type from different momenta and bands \cite{Weida}.
Note, the product of \emph{three} dipole matrix elements appearing in
$\chi^{(2)}_{\mathrm{irr}}$ is responsible for the surface sensitivity of
SFG, since in inversion symmetric crystals the product of dipole matrix
elements connecting three states vanishes except when inversion symmetry is
explicitly broken, \textit{e.g.}, by the surface.

In frequency space the nonlinear susceptibility is given by
\ba
\lefteqn{
\chi^{(2)}_{\mathrm{irr};ijk}(\q,\q_1,\q_2;\omega,\omega')
 = -\frac{e^3}{v} \sum_{\k_\|} \sum_{ll'l''} } \nonumber \\
& & {}\times
  \frac{D^i_{\k_\| l;\k_\| l''}(-q_z)}{-\hbar\omega
  +E_{\k_\| l}-E_{\k_\| l''}-i\hbar\Gamma_{\k_\| l'';\k_\| l}}
  \nonumber \\
& & {}\times
  \bigg[ D^j_{\k_\| l'';\k_\| l'}(q_{1z})\,
  D^k_{\k_\| l';\k_\| }(q_{2z})\,
  \nonumber \\
& & \qquad{}\times
  \frac{f(E_{\k_\| l})-f(E_{\k_\| l'})}
    {-\hbar\omega+\hbar\omega'+E_{\k_\| l}\!-E_{\k_\| l'}\!
      -i\hbar\Gamma_{\k_\| l';\k_\| l}}
  \nonumber \\
& & \quad{}- D^j_{\k_\| l';\k_\| l}(q_{1z})\,
  D^k_{\k_\| l'';\k_\| l'}(q_{2z})\,
  \nonumber \\
& & \qquad{}\times
  \frac{f(E_{\k_\| l'})-f(E_{\k_\| l''})}
    {-\hbar\omega+\hbar\omega'+E_{\k_\| l'}\!-E_{\k_\| l''}\!
      -i\hbar\Gamma_{\k_\| l'';\k_\| l'}}
  \bigg] .
\label{21.chiom}
\ea
This shows that the contribution of intermediate (virtual) states falls off
with the inverse of the initial-state energy plus the photon energy minus
the intermediate state energy, \textit{i.e.}, with the inverse of the
\emph{detuning}. This is not related to the lifetime broadening, but is due
to Heisenberg's uncertainty principle, which allows energy conservation to
be violated on short time scales. The frequency picture also allows to
incorporate a weight factor to account for the frequency resolution of the
detector \cite{Weida}. It is of interest to note that Eq.~(\ref{21.chiom})
and Fig.~\ref{fig.chi2inp} can also describe spin-selective electron
excitations due to circularly polarized light. Including the
electron spins our response theory and in particular Eq.~(\ref{21.chiom})
apply also to magnetic systems.

To prepare the analysis of the effect of collective plasma excitations on
the nonlinear optical response we now include the screening of the electric
fields. Screening enters in two ways: First, the effective field ${\bf E}$
within the solid is not identical to the external field because of linear
screening, which is expressed by the Fresnel formulae
\cite{SMD87,HBB94,Luce97,LuceLDA,AtK02} containing the dielectric
function $\varepsilon$, which can be determined in the RPA. Secondly, the
second-order polarization ${\bf P}^{(2)}$ of the electron gas, which
corresponds to a displacement of charge, leads to an additional electric
field \cite{Jackson}
\be
E_i^{(2)}(\r) = \int d^3r'\, \sum_j
  \left[\frac{3(r_i-r'_i)(r_j-r'_j)}{|\r-\r'|^5}
  - \frac{\delta_{ij}}{|\r-\r'|^3} - \frac{4\pi}{3}\,\delta_{ij}
  \,\delta(\r-\r')\right]\, P_j^{(2)}(\r') .
\label{21.Jackson2}
\ee
Fourier transformation leads to
\be
{\bf E}^{(2)}(\k,t) = -4\pi\,\hat\k\,\hat\k\cdot
  {\bf P}^{(2)}(\k,t) ,
\label{21.En1}
\ee
where $\hat\k$ is the unit vector in the direction of $\k$. Thus
only the component of ${\bf P}^{(2)}(\k,t)$ parallel to $\k$, \textit{i.e.},
its longitudinal part, is
accompanied by an electric field $\mathbf{E}^{(2)}$ \cite{Jackson},
which is also longitudinal. Note that
a longitudinal component of the electric field and of the induced
polarization generally exists even for a transverse applied external field
for lattice models \cite{Barton,Board}, see Eq.~(\ref{21.chiom}).

Due to the \emph{linear} polarizability of the solid the additional field
$\mathbf{E}^{(2)}$ leads to a polarization contribution of the form
$\chi\mathbf{E}^{(2)}$. Since the field $\mathbf{E}^{(2)}$ in
Eq.~(\ref{21.En1}) is of second order in the applied field, see
Eq.~(\ref{21.P2}), this polarization contribution must be taken into
account in $\mathbf{P}^{(2)}$. Doing this selfconsistently corresponds to
the summation of an RPA series \cite{HB}, as shown in App.~\ref{app.a}. Then, 
$P_i^{(2)}\propto \int dt_1\,dt_2\,\chi^{(2)}_{ijk}\,E_j\,E_k$ where now  
the nonlinear
susceptibility $\chi^{(2)}_{ijk}$ obtains an additional factor and
is given by
\ba
\lefteqn{
\chi^{(2)}_{ijk}(\q,\q_1,\q_2;t-t_1,t_1-t_2) = \frac{1}{2\pi}
  \sum_m \sum_{\tilde\q} \int d\tilde t
  } \nonumber \\
& & \quad{}\times
  \varepsilon_{\mathrm{long};im}^{-1}(\q,\tilde\q;t-\tilde t)\:
  \chi^{(2)}_{\mathrm{irr};mjk}(\tilde\q,\q_1,\q_2;\tilde t-t_1,t_1-t_2) .
\label{21.chiechi}
\ea
Here the irreducible susceptibility $\chi^{(2)}_{\mathrm{irr}}$ is given
by Eq.~(\ref{2.chi210}). Since only the longitudinal component of
$\mathbf{P}^{(2)}$ is accompanied by an electric field, the
screening factor $\varepsilon_{\mathrm{long}}^{-1}$ appears
only for the longitudinal component. This is expressed by the 
factor $\hat{\tilde q}_m \hat{\tilde q}_j$ in the explicit expression
$\varepsilon_{\mathrm{long};ij}(\q,\tilde\q;t-\tilde t)
  \equiv \delta_{ij} + 4\pi \chi_{im}(\q,\tilde\q;t-\tilde t)\,
  \hat{\tilde q}_m \hat{\tilde q}_j$.
\ignore{Note, $\varepsilon_{\mathrm{long}}$ is unity for the
transverse
components.}$\varepsilon_{\mathrm{long};im}^{-1}(\q,\tilde\q;t-\tilde t)$
is the inverse matrix with respect to the indices $(i,q_z)$ and
$(m,\tilde q_z)$.

\begin{figure}[ht]
\includegraphics[width=8.00cm]{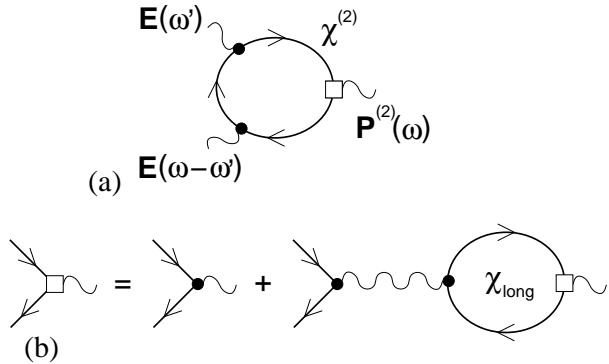}
\caption{(a) Diagrammatic representation of the second-order susceptibility
$\chi^{(2)}$ in terms of the effective electric field ${\bf E}$.
${\bf P}^{(2)}$ is the induced nonlinear polarization.
The square vertex $\Box$
represents the factor $\varepsilon_{\mathrm{long}}^{-1}$
at the frequency of ${\bf P}^{(2)}$. It appears if one self-consistently
takes into account the electric field due to the polarization
$\mathbf{P}^{(2)}$ of the electron
system and is given by the RPA series shown in Fig.~(b). The wriggly line
refers to the electron-electron interaction through the electromagnetic
field and is absorbed into the matrix elements $\mathbf{D}$. Note,
$\chi_{\mathrm{long}}$ only acts on the longitudinal field components, see
text. The factor $\varepsilon_{\mathrm{long}}^{-1}$ can be enhanced
by the plasma resonance.}
\label{fig.chi2}
\end{figure}

This analysis is illustrated by Fig.~\ref{fig.chi2}. Figure
\ref{fig.chi2}(a) shows the diagram of the second-order susceptibility
$\chi^{(2)}$. The square vertex represents the additional factor of
$\varepsilon_{\mathrm{long}}^{-1}$. It is obtained from the Dyson equation
in Fig.~\ref{fig.chi2}(b). The expression for $\chi^{(2)}_{\mathrm{irr}}$
in Eq.~(\ref{2.chi210}) is called irreducible since its diagram
Fig.~\ref{fig.chi2}(a) with the square vertex replaced by a normal one
cannot be cut into two by severing a single photon line.

The response theory clarifies how collective plasma excitations affect SFG.
They essentially enter in two ways, both of which are controlled by the
full (not only longitudinal) dielectric function $\varepsilon$:

First, the effective electric field is expressed in terms of the external
field by means of Fresnel formulae \cite{SMD87,HBB94,Luce97,LuceLDA,AtK02},
which contain contributions of order $1/\varepsilon$ for small
$\varepsilon$. The dielectric function $\varepsilon$ becomes small if the
frequency of the external field is close to the plasma frequency. This
contribution can be interpreted as \emph{field enhancement}. In addition,
the outgoing (sum-frequency) electric field ${\bf E}_{\mathrm{out}}$ also
contains terms that are enhanced for small $\varepsilon$ due to the Fresnel
factors. This enhancement is most pronounced if the sum frequency is close
to the plasma frequency.

Secondly, the longitudinal component of the nonlinear polarization ${\bf
P}^{(2)}$ of the electron system is accompanied by an electric field ${\bf
E}^{(2)}$ given by Eq.~(\ref{21.En1}). Thus, the factor
$\varepsilon_{\mathrm{long}}^{-1}$ appears in the nonlinear susceptibility
in Eq.~(\ref{21.chiechi}) and thus in $\mathbf{P}^{(2)}$ \cite{HB}. This
leads to an enhancement of the SFG light due to the longitudinal part of
$\mathbf{P}^{(2)}$ if the sum frequency is close to the plasma frequency.

\subsection{Two-photon photoemission}
\label{ss.22}

To demonstrate the similarities between SFG and 2PPE, we continue by
summarizing the results of the response theory for 2PPE. We consider the
same band structure as for SFG, which is characterized by single-electron
energies $E_{\k_\| l}$. We emphasize that this band structure contains the
bulk states with the $z$-component $k_z$ of $\k$ included in $l$.

The response theory starts from the observation that the photoelectron
current $j(t;{\bf k},\sigma)$ of electrons of momentum $\k$ and spin
$\sigma$ is given by the change of occupation of the vacuum state
$|\k\sigma,\ou\rangle$ outside of the crystal. However, in practice the
time-dependence of $j$ is not measured, but only the total photoelectron
yield ${\cal N}(\k,\sigma) = \int dt\: j(t;{\bf k},\sigma)$. This is
similar to SFG, where only the time-integrated intensity is measured. The
response theory directly determines the photoelectron yield ${\cal N}$. To
prepare the discussion it is useful to first consider ordinary
single-photon photoemission.

\begin{figure}[ht]
\includegraphics[width=4.00cm]{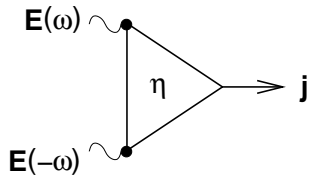}
\caption{Diagrammatic representation of the relation $\mathbf{j}\propto \int
dt_1\,dt_2\, \eta_{ij}\, E_i E_j$ for the photoelectron current in ordinary
photo\-emis\-sion \protect\cite{Feder}.
The wavy lines denote the effective electric field ${\bf E}$
within the solid and the arrow denotes the emitted electron current ${\bf j}$
(or the photoelectron yield), which is of second order in the electric field.
The response function $\eta$ is discussed in the text.
The dots ($\bullet$) denote dipole matrix elements $\mathbf{D}$.}
\label{fig.N2}
\end{figure}

\textbf{Single-photon photoemission:} The photoelectron yield is given by
\be
{\cal N}(\k,\sigma) = \sum_{\q}
  \int dt_1\,dt_2\,
  \eta_{ij}(\q;t_1,t_2;\k,\sigma)\, E_i(\q,t_1)\,
  E_j(-\q,t_2) ,
\label{2.N2.1}
\ee
with the response function (see App.~\ref{app.b})
\ba
\lefteqn{
\eta_{ij}(\q;t_1,t_2;\k,\sigma)
  = \frac{e^2}{\hbar^2}
  \frac{\gamma_{\k\sigma,\ou;\k\sigma,\iv}}
    {\Gamma_{\k\sigma,\iv;\k\sigma,\iv}}
  \sum_\lambda
   D^i_{\k\sigma,\iv;\k_\| \lambda}(q_z) } \nonumber \\
& & \quad{}\times
   \exp\!\left[i\frac{E_{\k_\| \lambda}-E_{\k\sigma,\iv}}{\hbar}
     (t_2-t_1)\right]
   e^{-\Gamma_{\k_\| \lambda;\k\sigma,\iv}|t_2-t_1|}
  \nonumber \\
& & \quad{}\times
   f(E_{\k_\| \lambda})\,
   D^j_{\k_\| \lambda;\k\sigma,\iv}(-q_z) .
   \qquad\qquad\qquad\quad\;\;\;
\label{2.eta2.1}
\ea
Here, $|\k\sigma,\iv\rangle$ is a state with momentum $\k$ and spin
$\sigma$ inside the crystal but above the vacuum energy. We have again
neglected the momentum transferred by the photon. The standard diagrammatic
representation of ordinary photoemission is shown in Fig.~\ref{fig.N2}
\cite{Feder}. The effective field ${\bf E}$ within the solid should again
be expressed in terms of the external light field with the help of the
proper boundary conditions. The response function $\eta$ will play a role
when we discuss the various contributions to 2PPE.

We briefly commend on the structure of this expression: The prefactor
$\gamma_{\k\sigma,\ou;\k\sigma,\iv}/\Gamma_{\k\sigma,\iv;\k\sigma,\iv}$
describes the probability that electrons excited above the vacuum energy
actually leave the crystal. Photoemission is often described by a
three-step picture \cite{BS,FFW,WD}: First, electrons are excited, then they
are transported to the surface, and finally they leave the crystal. In this
work we are mainly interested in the first step. The second and third steps
are incorporated phenomenologically by effective relaxation rates
$\Gamma_{\k\sigma,\iv;\k\sigma,\iv}$, which describe electrons dropping
below $\Ev$ before they reach the surface, and effective transition rates
$\gamma_{\k\sigma,\ou;\k\sigma,\iv}$ from states above $\Ev$ within the
solid to free electron states outside of the solid.
Note, the yield is proportional to the electric field squared and thus to
the intensity of the incoming light.

\begin{figure}[ht]
\includegraphics[width=8.50cm]{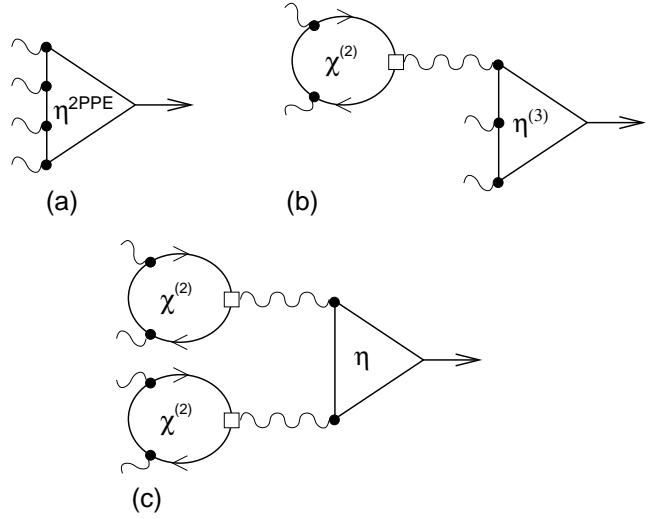}
\caption{Diagrammatic representation of contributions to the
2PPE yield ${\cal N}^{\mathrm{2PPE}}$. (a) Direct
irreducible contribution ${\cal N}^{\mathrm{2PPE}}_{\mathrm{irr}}$
involving four transitions induced by the
effective electric field ${\bf E}$.
(b) Reducible process involving conversion of two photons to a SFG photon
and subsequent photoemission of third order in the fields, yielding ${\cal
N}^{\mathrm{2PPE}}_{\mathrm{red,1}}$.
(c) Reducible process involving conversion of all four photons to two
SFG photons and ordinary photoemission (of second order in the fields)
induced by the SFG light, yielding ${\cal
N}^{\mathrm{2PPE}}_{\mathrm{red,2}}$. The dots ($\bullet$) denote dipole
matrix elements $\mathbf{D}$. The square vertex ($\Box$),
representing a factor of $\varepsilon_{\mathrm{long}}^{-1}$, is defined
in Fig.~\protect\ref{fig.chi2}(b). Plasma excitations again enter through
this vertex.}
\label{fig.N4}
\end{figure}

\textbf{Two-photon photoemission:}
The total 2PPE yield consists of the three contributions
\be
{\cal N}^{\mathrm{2PPE}} = {\cal N}^{\mathrm{2PPE}}_{\mathrm{irr}}
  + {\cal N}^{\mathrm{2PPE}}_{\mathrm{red,1}}
  + {\cal N}^{\mathrm{2PPE}}_{\mathrm{red,2}}
\label{22.N2tot}
\ee
corresponding to Fig.~\ref{fig.N4}(a), (b), and (c), respectively.
The second and third term arise from the nonlinear optical properties of
the solid: Close to the surface
the effective field ${\bf E}$ leads to a second-order
polarization ${\bf P}^{(2)}$, see Eq.~(\ref{21.P2}), which is accompanied by
an electric field ${\bf E}^{(2)}$. This field may contribute to
photoemission, leading to the processes in Fig.~\ref{fig.N4}(b) and (c).
The second-order field ${\bf E}^{(2)}$ is also responsible for SFG
accompanying 2PPE. However, this SFG is usually a small effect since the SFG
light \emph{intensity} is of \emph{sixth} order in dipole matrix elements
$\mathbf{D}$, see Eqs.~(\ref{21.S2.1}) and (\ref{2.chi210}), whereas the
2PPE current is of \emph{fourth} order, as is shown below in
Eq.~(\ref{22.eta4.sh}). This changes if the sum frequency is
close to the plasma frequency, in which case SFG is enhanced as discussed
at the end of Sec.~\ref{ss.21}.

Since the diagram in Fig.~\ref{fig.N4}(a) cannot be cut into two by
severing a single photon line, the first term
${\cal N}^{\mathrm{2PPE}}_{\mathrm{irr}}$
is irreducible, while the other two are reducible. The irreducible
contribution in Eq.~(\ref{22.N2tot}) can be written as
\be
{\cal N}^{\mathrm{2PPE}}_{\mathrm{irr}}
  = \int dt_1\,dt_2\,dt_3\,dt_4\;
  \eta^{\mathrm{2PPE}}_{ijkl}(t_1,t_2,t_3,t_4)\,
  E_i(t_1)\,E_j(t_2)\,E_k(t_3)\,E_l(t_4) ,
\label{22.N2irr}
\ee
which is of \emph{fourth} order in the electric field and of \emph{second}
order in the incoming intensity. This is already clear from the simple
picture in Fig.~\ref{fig2}: The occupation of the intermediate state
$|2\rangle$ is proportional to the light intensity. To reach state
$|3\rangle$ above $\Ev$ another absorption is required, leading to a total
proportionality to the intensity squared.

Obviously, the structure of Eq.~(\ref{22.N2irr}) is very similar to
Eq.~(\ref{21.S2.1}) for the SFG yield:
\ba
{\cal I}^{(2)} & \propto & \int dt\,dt_1\,dt_2\,dt_3\,dt_4\;
  \chi^{(2)}_{ijk}(t-t_1,t_1-t_2)\,\chi^{(2)}_{ilm}(t-t_3,t_3-t_4)
  \nonumber \\
& & {}\times E_j(t_1)\,E_k(t_2)\,E_l(t_3)\,E_m(t_4) .
\ea
Hence, we expect similar interference effects in both cases.

The other two contributions to ${\cal N}^{\mathrm{2PPE}}$ are
\ba
\lefteqn{
{\cal N}^{\mathrm{2PPE}}_{\mathrm{red,1}}
  = -4\pi \int dt_1\,dt_2\,dt_3\;
    \eta^{(3)}_{ijk} \: \Big[ P_i^{(2)}(t_1)\,E_j(t_2)\,E_k(t_3)
  } \nonumber \\
& & \; {}+ E_i(t_1)\,P_j^{(2)}(t_2)\,E_k(t_3)
  + E_i(t_1)\,E_j(t_2)\,P_k^{(2)}(t_3) \Big] ,
\label{22.N2red1}
\ea
and
\be
{\cal N}^{\mathrm{2PPE}}_{\mathrm{red,2}} = (4\pi)^2 \int dt_1\,dt_2\;
    \eta_{ij} \: P_i^{(2)}(t_1)\,P_j^{(2)}(t_2) ,
\label{22.N2red2}
\ee
where $\eta$ is given in Eq.~(\ref{2.eta2.1}). Since the nonlinear
susceptibility $\chi^{(2)}$ and hence $\mathbf{P}^{(2)}$ contains three
dipole matrix elements, the reducible contributions to the photoelectron
current are of higher order in dipole matrix elements and are thus usually
small. However, the longitudinal component of ${\bf P}^{(2)}$ contains a
factor $\varepsilon_{\mathrm{long}}^{-1}$. If the nonlinear polarization
is enhanced due to a bulk plasma resonance at the sum frequency, one
expects significant contributions from the reducible terms. The response
functions $\eta^{(3)}$ and $\eta^{\mathrm{2PPE}}$ are given in
App.~\ref{app.b}.

We next consider the response functions $\eta$, $\eta^{(3)}$, and
$\eta^{\mathrm{2PPE}}$ which determine the yield ${\cal
N}^{\mathrm{2PPE}}$. The functions $\eta^{(3)}$ and $\eta^{\mathrm{2PPE}}$ 
appearing in Eqs.~(\ref{22.N2red1}) and (\ref{22.N2red2}), respectively,
are of the same general form as $\eta$, Eq.~(\ref{2.eta2.1}), but have more
terms resulting from different orders of the time arguments. We now first
present the structure of the response expression for the main, irreducible
contribution to the 2PPE yield, Eq.~(\ref{22.N2irr}), and then discuss its
physical interpretation. Fully written out, Eq.~(\ref{22.N2irr}) reads
\ba
{\cal N}_{\mathrm{irr}}^{\mathrm{2PPE}}(\k,\sigma)
  & = & \sum_{\q_1\q_2\q_3} \int dt_1\,dt_2\,dt_3\,dt_4\:
  \eta^{\mathrm{2PPE}}_{ijkl}(\q_1,\q_2,\q_3;t_1,t_2,t_3,t_4;\k,\sigma)\,
  E_i(\q_1,t_1)\,E_j(\q_2,t_2)\,
  \nonumber \\
& & {}\times
  E_k(\q_3,t_3)\,E_l(-\q-\q_1-\q_2,t_4) .
\label{22.N2irrf}
\ea
Defining the complex transition energy
\be
\Omega_{\k_\| l;\k'_\| l'} \equiv \frac{E_{\k_\| l}-E_{\k'_\| l'}}{\hbar}
  - i\Gamma_{\k_\| l;\k'_\| l'} ,
\label{22.Om1}
\ee
we obtain the response function
\ba
\lefteqn{
\eta^{\mathrm{2PPE}}_{ijkl}(\q_1,\q_2,\q_3;t_1,t_2,t_3,t_4;\k,\sigma)
  = \frac{e^4}{\hbar^4}\,
    \frac{\gamma_{\k\sigma,\ou;\k\sigma,\iv}}
    {\Gamma_{\k\sigma,\iv;\k\sigma,\iv}}
  \sum_{\lambda_1\lambda_2\lambda_3}
  D^i_{\k\sigma,\iv;\k_\| \lambda_1}(q_{1z})\,
  D^j_{\k_\| \lambda_1;\k_\| \lambda_2}(q_{2z})\,
  } \nonumber \\
& & \; {}\times
  D^k_{\k_\| \lambda_2; \k_\| \lambda_3}(q_{3z})\,
  D^l_{\k_\| \lambda_3; \k\sigma,\iv}(-q_{1z}-q_{2z}-q_{3z})
  \nonumber \\
& & \; {}\times
  \bigg\{
  \Theta(t_1-t_2)\,\Theta(t_2-t_3)\,\Theta(t_3-t_4)\,
  e^{-i\Omega_{\k_\| \lambda_1;\k\sigma,\iv}(t_1-t_2)}\,
  e^{-i\Omega_{\k_\| \lambda_2;\k\sigma,\iv}(t_2-t_3)}\,
  e^{-i\Omega_{\k_\| \lambda_3;\k\sigma,\iv}(t_3-t_4)}
  \left[-f(E_{\k_\| \lambda_3})\right]
  \ignore{\!\!\!\!\!\!\!\!\!\!\!\!\!\!\!\!\!\!\!\!\!\!\!\!\!\!\!}
  \nonumber \\
& & \quad {}-\Theta(t_1-t_2)\,\Theta(t_2-t_4)\,\Theta(t_4-t_3)\,
  e^{-i\Omega_{\k_\| \lambda_1;\k\sigma,\iv}(t_1-t_2)}\,
  e^{-i\Omega_{\k_\| \lambda_2;\k\sigma,\iv}(t_2-t_4)}\,
  e^{-i\Omega_{\k_\| \lambda_2;\k_\| \lambda_3}(t_4-t_3)}
  \left[f(E_{\k_\| \lambda_3})\!-\!f(E_{\k_\| \lambda_2})\right]
  \ignore{\!\!\!\!\!\!\!\!\!\!\!\!\!\!\!\!\!\!\!\!\!\!\!\!\!\!\!}
  \nonumber \\
& & \quad {}- \ldots\bigg\} .
\label{22.eta4.sh}
\ea
There are eight terms in the curly braces, which correspond to different
temporal orders of interactions with the electric field. Note, the
dependence of 2PPE on light polarization is incorporated in the symmetries
of the tensor $\eta^{\mathrm{2PPE}}_{ijkl}$, which depend on the dipole
matrix elements $\mathbf{D}$. Unlike for the nonlinear optical response,
these symmetries have not been discussed so far. It would be very
interesting to determine the symmetries for surfaces of nonmagnetic and
magnetic solids.

\begin{figure}[ht]
\includegraphics[width=6.00cm]{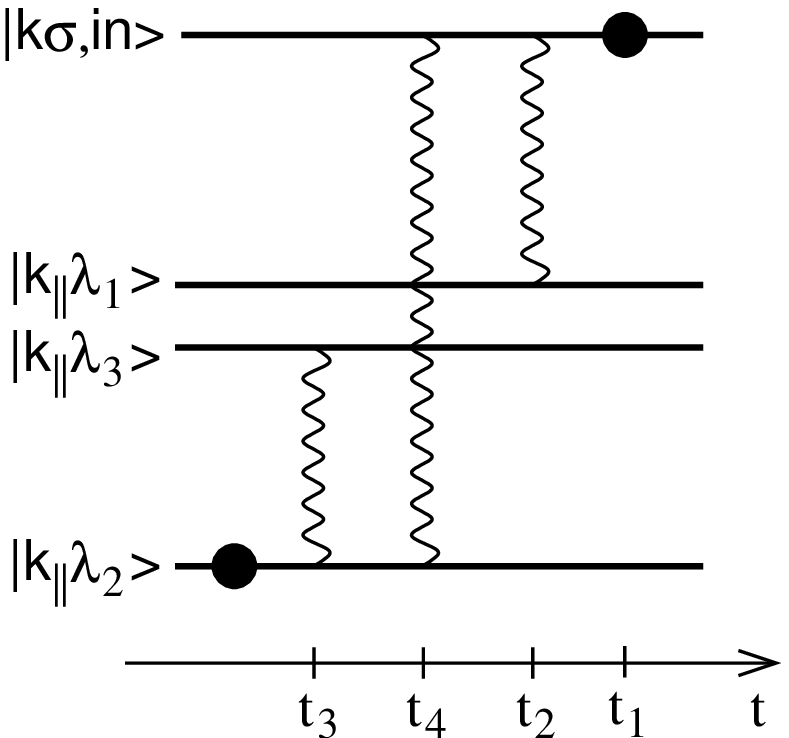}
\caption{Interpretation of one of the processes contributing to 2PPE.
An electron is excited from a pure state $|\k_\|\lambda_2\rangle$
in the Fermi sea to a pure state $|\k\sigma,\iv\rangle$ above
$\Ev$ by four interactions with the electric field at
the times $t_i$, as expressed by Eqs.~(\protect\ref{22.N2irrf}) and
(\protect\ref{22.eta4.sh}).
Note, the photoelectron current is proportional to the
fourth power of the electric field and thus to the intensity \emph{squared},
as expected for \emph{two-photon} photoemission.
The heavy wavy lines denote superpositions of states
$|\k_\|\lambda_2\rangle$ and $|\k_\|\lambda_3\rangle$,
$|\k_\|\lambda_2\rangle$ and $|\k\sigma,\iv\rangle$ etc.,
while the black
dots represent pure eigenstates. Compare with Fig.~\protect\ref{fig.chi2inp}
for SFG.}
\label{fig.N4inp}
\end{figure}

Equation (\ref{22.eta4.sh}) forms the basis for our discussion of 2PPE. To
clarify the time dependence exhibited in Eq.~(\ref{22.eta4.sh}) we discuss
the second term, the others are in principle similar but correspond to
different orders of the times $t_i$. The processes described by this term
are illustrated in Fig.~\ref{fig.N4inp}. The system starts in equilibrium
from the state $|\k_\|\lambda_2\rangle$. The first interaction with the
electric field takes place at time $t_3$ and creates a superposition of the
states $|\k_\|\lambda_2\rangle$ and $|\k_\|\lambda_3\rangle$, leading to
oscillations at the frequency
$(E_{\k_\|\lambda_3}-E_{\k_\|\lambda_2})/\hbar$ expressed by the third
exponential factor in this term. The Fermi factors ensure that one of the
states is initially occupied and that the other one is empty. Let us assume
that $|\k_\|\lambda_2\rangle$ is occupied. The second interaction at $t_4$
changes the state into a superposition of $|\k_\|\lambda_2\rangle$ and the
vacuum state $|\k\sigma,\iv\rangle$, leading to oscillations at the
corresponding difference frequency (second exponential factor), and the
third interaction at $t_2$ creates a superposition of the vacuum state and
$|\k_\|\lambda_1\rangle$. After the fourth interaction the electron is in a
pure state above $\Ev$ and can leave the solid with finite
probability. Of course, due to the sum over bands there are usually several
contributions of this type. Only if the superpositions decay very rapidly
compared to the pure states, a description in terms of rate equations, as
suggested by Fig.~\ref{fig2}, is applicable \cite{Kno00,Rol1,Rol2}. Also
compare the discussion of SFG above, see Fig.~\ref{fig.chi2inp}.

While SFG is only governed by the dephasing rates but not the energy
relaxation rates, 2PPE depends on both. This is because in the 2PPE
response function $\eta^{\mathrm{2PPE}}$ the change of occupation of states enters
besides the polarization of the electron gas, whereas SFG only depends on
the latter.

Note, the 2PPE yield contains {\it four\/} dipole matrix elements. Thus,
even for inversion-symmetric crystals parity does not forbid 2PPE from the
bulk. However, 2PPE is sensitive to a surface region of a thickness given
by the mean free path of electrons above $\Ev$. The optical penetration
depth is typically significantly larger than the mean free path and thus
does not enter here. Equations (\ref{2.eta2.1}) and (\ref{22.eta4.sh}) also
illustrate that 2PPE is sensitive to specific points in the Brillouin zone:
The photoelectron momentum $\k$ measured by {\it momentum-resolved\/} 2PPE
is approximately the same as the lattice momentum of the original
unperturbed electron and also of the intermediate state due to the small
photon momentum. These effects obviously require a theoretical description
that considers the $\k$-dependent states in the solid, like our approach
does as opposed to both the random-$\k$ approximation and Bloch equations.
In view of the importance of angle-resolved (ordinary) photoemission
spectroscopy (ARPES) for, {\it e.g.}, cuprate high-$T_c$ superconductors,
$\k$-resolved 2PPE is expected to yield interesting results in the future.
On the other hand, if one only measures the total number of photoelectrons,
the $\k$-space resolution is lost and 2PPE and SFG give very similar
information. It is obvious that 2PPE has the disadvantage of being limited
to frequencies $\omega_1$, $\omega_2$ such that
$E_F+\hbar\omega_1+\hbar\omega_2$ lies above the vacuum energy,
unlike SFG. As stated already the SFG accompanying 2PPE is usually small
since additional dipole matrix elements are involved.

The response expressions show that collective plasma excitations affect
2PPE in two ways: First, exactly like for SFG the effective field ${\bf E}$
within the metal differs from the external field due to linear screening
and is enhanced close to the plasmon resonance. Secondly, the reducible
contributions in Eqs.~(\ref{22.N2red1}) and (\ref{22.N2red2}) depend on the
second-order polarization ${\bf P}^{(2)}$, which contains a factor
of $\varepsilon_{\mathrm{long}}^{-1}$, see Eq.~(\ref{21.chiechi}). ${\bf
P}^{(2)}$ is enhanced if the sum frequency is close to the plasma
frequency. In 2PPE this enhancement enters only in the reducible
contributions in Eqs.~(\ref{22.N2red1}) and (\ref{22.N2red2}).

\section{Discussion}
\label{sec.disc}

The aim of the present section is to discuss and illustrate the results of
the response theory for time-resolved SFG and 2PPE. In particular, we
consider time-dependent effects on the femtosecond time scale. Our results
exhibit the intimate relation between SFG and 2PPE. We can already gain
insight by studying the general structure of the response expressions for
SFG and 2PPE, for example Eqs.~(\ref{21.S2.1}) and (\ref{22.N2irr}),
respectively, independently of the specific approximations made here. For
clarity we apply our response theory to a simple model system.

\subsection{Time-dependent effects in SFG and 2PPE}

The response expressions of the preceding section are valid for any time
dependence of the exciting laser field. The time enters the response
expressions for both SFG and 2PPE in two ways, apart from the step
functions from causality, cf.\ Eqs.~(\ref{2.chi210}), (\ref{2.eta2.1}), and
(\ref{22.eta4.sh}): The difference between the time arguments of electric
fields appears in exponentials oscillating at the transition frequency of
the involved electron states and in exponentials decaying with the
dephasing rate of the superposition of the two states, and, for 2PPE, also
exponentials decaying with the energy relaxation rate of an intermediate
state. (See the discussion of Figs.~\ref{fig.chi2inp} and \ref{fig.N4inp}
for the interpretation of SFG and 2PPE in terms of electronic excitations.)
The time passing between absorptions can be controlled by the pulse shape
of the exciting laser pulses: If the total duration $T$ of a pulse of
arbitrary shape is much larger than typical relaxation times $\tau$ then
the yield depends on the probability to absorb two photons within a time
interval $\tau$, which is independent of $T$. On the other hand, for
$T\ll\tau$ there is almost no relaxation during the pulse. Thus the
response theory reproduces the well-known result that $\tau$ can only be
inferred from SFG or 2PPE experiments if the total pulse duration is
$T\sim\tau$.

To be more specific, in most experiments two approximately Gaussian pulses
are used (pump-probe method)
\cite{Ste92,HMKB97,Sim98,KSR98,Gud99,Schoe,Aesch1,Aesch2,Wolf1,%
Hoef97,Cao97,Petek,Wolf2,Leh99,Kno00,GWH00,SBW02,ONP02,Gerb,Ae.cluster}. If
the two pulses are of different mean frequencies $\omega_1$ and $\omega_2$
(two-color case) and one measures the SFG or 2PPE response at the sum
frequency $\omega_1+\omega_2$ then it is obvious which photon was absorbed
out of which pulse. Then for long time delay $\Delta T$ compared to the
single-pulse duration the relaxation rate of intermediate states can be
read of directly from the $\Delta T$ dependence of the total yield. In
pump-probe experiments with two pulses of the same mean frequency $\omega$
(single-color case), photons can be absorbed out of the same or different
pulses. However, the contribution with all absorptions out of the same
pulse obviously does not depend on $\Delta T$, just leading to a constant
background. Note, in all these cases only a {\it typical\/} relaxation time
enters, which usually is a weighted average over relaxation times of many
states \cite{Weida}. If only a single relevant intermediate state is
present, {\it e.g.}, for a quantum-well state, or if there are many but of
similar relaxation rate, the relaxation time extracted from experiment will
be the actual dephasing time of intermediate states. However, if
intermediate states with very different dynamical properties are involved,
for example if both {\it sp\/} and {\it d\/} bands are relevant, the
measured relaxation time does not describe any single excited electron
state.

In pump-probe SHG \cite{Ste92,HMKB97,Sim98,KSR98,Gud99} or pump-probe
single-color 2PPE \cite{Aesch2,Cao97,Petek,Wolf2,Kno00} experiments,
time-dependent interference effects are especially pronounced. Their origin
is the following: The first absorption of a photon of frequency $\omega$
sets up an oscillating polarization of the excited electrons. Now the
probability of a second absorption depends on the relative phase of the
oscillating polarization and the second photon. Since the oscillating
polarization is described by the {\it off-diagonal\/} components of the
density matrix $\rho$, a description in terms of rate equations, which
omits these components, is unable to describe interference.

For further illustration of this interference, we show results for SHG and
2PPE for a simple model. Unless stated otherwise, this model consists of
three bands. The lowest one is a three-dimensional tight-binding band $1$
with band center at \cite{rem.param} $-3.33\:{\rm eV}$ (all energies are
measured relative to the Fermi energy) and half width $3.81\:{\rm eV}$. The
band maximum is at $\k=0$. The second, rather flat tight-binding band $2$
is centered at $2.29\:{\rm eV}$ with half width $0.48\:{\rm eV}$ and
maximum also at $\k=0$. Finally, there is a free electron band $3$
representing electrons above the vacuum energy $\Ev=4.29\:{\rm eV}$. There
exist points in the Brillouin zone for which the energy differences between
bands 2 and 1 as well as between bands 3 and 2 both equal the photon energy
of $\hbar\omega=3.05\:{\rm eV}$. We assume that the relaxation rates
$\Gamma_{n_1n_2}$ only depend on the band indices $n_1$, $n_2$ but not on
the $\k$ vector (see below). We use the energy relaxation rates
$\hbar\Gamma_{22}=0.191\:{\rm eV}$ (corresponding to the lifetime
$\tau_2=3.5\:{\rm fs}$) and $\hbar\Gamma_{33}=0.381\:{\rm eV}$
($\tau_3=1.7\:{\rm fs}$) and no additional dephasing, {\it i.e.},
$\Gamma_{n_1,n_2}^{\mathrm{ph}}=0$ in Eq.~(\ref{2.rates1}). These short
lifetimes are assumed to bring out the time-dependent effects more clearly.
The dipole matrix elements are treated as constants.

In the following we use this model to show how time-dependent effects
emerge from our response theory. For clarity we neglect the Fresnel
formulae, which do not change the results qualitatively. We demonstrate
that our theory gives reasonable results for a moderately complicated
system. Obviously, it can be applied to a more realistic band structure at
the expense of computation time. The boundary conditions (Fresnel factors)
are also omitted for simplicity.

\begin{figure}[ht]
\includegraphics[width=9.00cm]{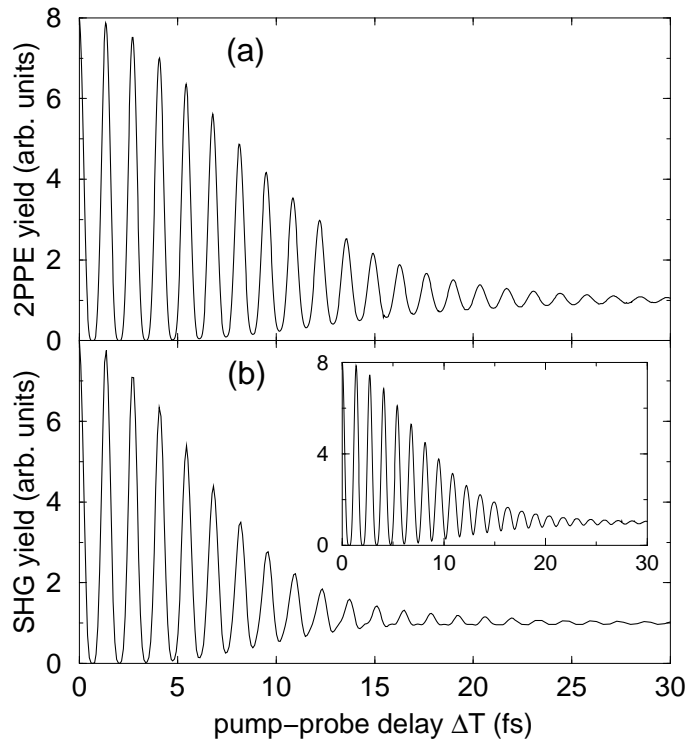}
\caption{(a) Total yield of photoelectrons of momentum $\k$ for
single-color pump-probe 2PPE as a function of the delay time $\Delta T$
between pump and probe pulses.
The parameters of the model are given in the text. The $\k$ vector is
chosen such that the transition frequencies perfectly match the frequency
of incoming light. Only results for $\Delta T>0$ are shown, since the curve
is symmetric about $\Delta T=0$ for identical pump and probe pulses. All
curves in this and the following figures are scaled such that the limit for
large $\Delta T$ is unity. (b) Total SHG yield for single-color pump-probe
SHG as a function of the delay time $\Delta T$ using the same parameters.
The inset shows the SHG yield for \emph{flat} bands with transition
frequencies that match the light frequency perfectly. Note, for extracting
relaxation rates from experimental data the resolution of the photoelectron
and SHG light detectors must be taken into account \protect\cite{Weida}.}
\label{fig.interPS}
\end{figure}

In Fig.~\ref{fig.interPS}(a) we show the 2PPE photoelectron yield for a
particular momentum $\k$ as a function of the delay time $\Delta T$ between
two identical Gaussian pump and probe pulses. A mean photon energy of
$\hbar\omega=3.05\:{\rm eV}$ is assumed, corresponding to a wave length of
about $\lambda=400\:{\rm nm}$, and the duration of each pulse is
$10.3\:{\rm fs}$ (full width at half maximum of the Gaussian envelope of
the electric field). The vector $\k$ is chosen so that the transition
energies between the bands match $\hbar\omega$. In
Fig.~\ref{fig.interPS}(b) we show the total SHG photon yield for exactly
the same system. Unlike 2PPE, SHG integrates over the whole Brillouin zone.
Nevertheless, the overall similarity of Figs.~\ref{fig.interPS}(a) and (b)
demonstrates the similarity of the response expressions for SHG and 2PPE,
compare Eqs.~(\ref{21.S2.1}) and (\ref{22.N2irr}), for example. It means
that similar information, {\it e.g.}, about the relaxation rates, can be
obtained from both. The SHG curve is quite similar to the case of flat
bands, shown in the inset in Fig.~\ref{fig.interPS}(b). This means that
only a small region of $\k$ space contributes. The resulting interference
between different $\k$ points becomes apparent in the tail of the
interference pattern, where the main plot in Fig.~\ref{fig.interPS}(b) is
more irregular and decays faster. This is the averaging effect discussed in
Ref.~\cite{Weida}. More precisely it is an \emph{interference} effect
between different oscillation frequencies of superpositions of different
states.

The 2PPE and SHG interference patterns in Fig.~\ref{fig.interPS}
show the well-known $8:1$ enhancement of
the signal for $\Delta T=0$. This enhancement is due to the yield
being of fourth order in the field: For a single pulse the signal would be
proportional to $E^4$, for two isolated pulses this becomes $2E^4$, but for
two overlapping pulses the amplitude is doubled, leading to
$(2E)^4 = 16 E^4$.

\begin{figure}[ht]
\includegraphics[width=9.00cm]{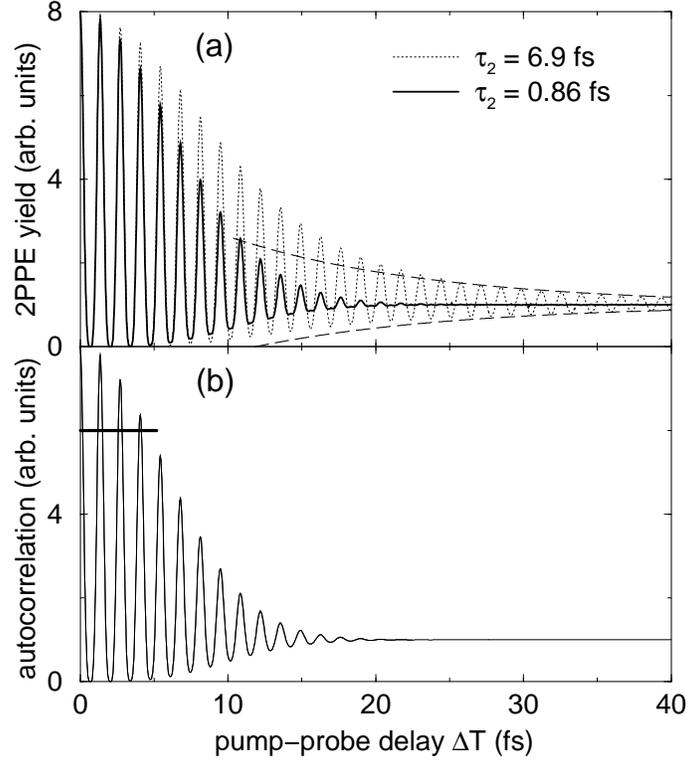}
\caption{Demonstration of the lifetime dependence of 2PPE.
(a) Total 2PPE yield for the same model parameters as used in
Fig.~\protect\ref{fig.interPS} with a lifetime of states in the intermediate
band of $\tau_2=6.9\:{\rm fs}$ (dotted curve) and with the very small
value $\tau_2=0.86\:{\rm fs}$ (heavy solid curve). The dashed curves show
the exponential decay with the dephasing rate $\Gamma_{12}=\tau_2^{-1}/2$
for $\tau_2=6.9\:{\rm fs}$.
(b) Four-field autocorrelation function of the pump-probe laser field.
Note the similarity to the fast-relaxation result in Fig.~(a).
The black bar denotes half the laser pulse duration.}
\label{fig.iP.relax}
\end{figure}

For both SHG and 2PPE, the central part of the interference pattern,
which corresponds to short delay
times $\Delta T$ up to about the single-pulse duration $T$, is dominated by
the four-field autocorrelation function
\be
{\cal A}_{ijkl}^{(4){}}(\Delta T) \equiv \int d\omega_1\,d\omega_2\,
  d\omega_3\; E_i(\omega_1)\,E_j(\omega_2)\,E_k(\omega_3)
  E_l(-\omega_1-\omega_2-\omega_3) .
\ee
This central part stems from the overlap of
the two pulses and would be present even for very fast relaxation: Then
the response functions $\chi^{(2)}$ and $\eta^{\mathrm{2PPE}}$ are very sharply peaked
in time and thus nearly constant in frequency space, leading to
${\cal I}^{(2)} \propto {\cal A}^{(4){}}$ and
${\cal N}^{\mathrm{2PPE}}_{\mathrm{irr}} \propto {\cal A}^{(4){}}$
for the SHG and 2PPE yield, respectively, see Eqs.~(\ref{21.S2.1})
and (\ref{22.N2irr}). The autocorrelation signal alone is shown
in Fig.~\ref{fig.iP.relax}(b).

In Sec.~\ref{sec.theory} we have discussed the response expressions for
time-dependent SFG and 2PPE, (\ref{2.chi210}) and (\ref{22.eta4.sh}),
respectively. The first interaction creates an oscillating polarization.
There is interference if the phase information is still preserved when the
second photon is absorbed. This is governed by the dephasing time
$\Gamma_{21}$. Thus the interference effects should decay with the time
constant $\Gamma_{21}^{-1}$ for large delays $\Delta T$. This is shown in
Fig.~\ref{fig.iP.relax}(a) for moderately fast ($\tau_2=6.9\:{\rm fs}$) and
extremely fast ($\tau_2=0.86\:{\rm fs}$) relaxation. For the slower
relaxation the tail indeed decays with $\Gamma_{21}^{-1}$ but to observe
this one obviously has to look at rather large $\Delta T$ where the
interference is already weak. For fast relaxation the curve is nearly
indistinguishable from the autocorrelation in Fig.~\ref{fig.iP.relax}(b).

\begin{figure}[ht]
\includegraphics[width=9.00cm]{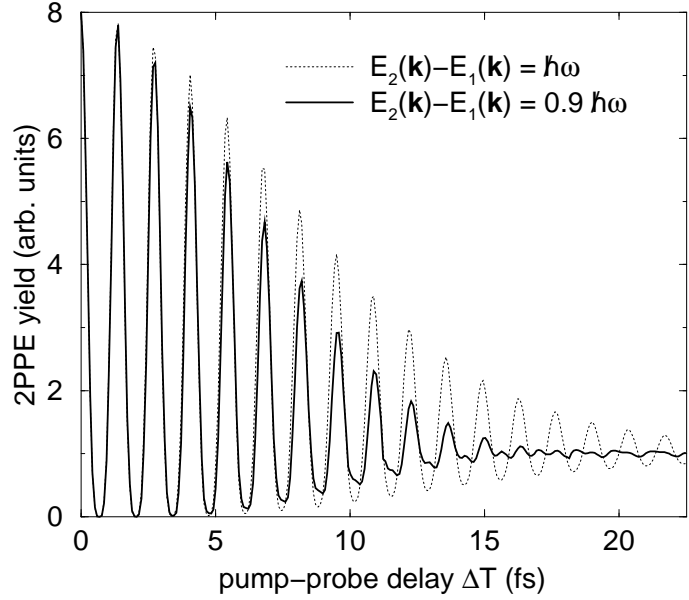}
\caption{Demonstration of the dependence of 2PPE on the detuning of the
intermediate band. The heavy solid curve shows the total 2PPE yield for the
same parameters as used in Fig.~\protect\ref{fig.interPS} but with the
intermediate band shifted downward in energy by $0.1\,\hbar\omega$.
$\omega$ is the mean frequency of the exciting laser field. For comparison,
the dotted curve shows the 2PPE yield for unshifted bands with matching
transition frequencies. Note the beats apparent for $\Delta T \gtrsim
10\:\mathrm{fs}$. This shows how
detuning affects the decay of the signal and needs to be taken into account
when extracting relaxation rates from experiments.}
\label{fig.iP.det}
\end{figure}

However, there is another crucial origin of the decay of interference:
Intermediate states with energies that do not exactly match the energy of
the original state plus the photon energy lead to {\it beats\/} at the
frequency of the detuning. This effect can be seen from the response
expressions. We now discuss this for the case of SFG: For pulses of short
duration $T$, the times $t_1$ and $t_2$ in Eq.~(\ref{2.chi210}) can
be approximated by the pulse centers if we are interested in phenomena
at frequencies small compared to $T^{-1}$. Then Eq.~(\ref{2.chi210}) shows
that the polarization of the electron system shortly before the second
interaction at $t_1$ is proportional to
\be
\exp\!\left[i\frac{E_{\k_\|l}-E_{\k_\|l'}}{\hbar}\,(t_1-t_2)\right]
  e^{-\Gamma_{\k_\|l';\k_\|l}(t_1-t_2)} e^{-i\omega t_2} ,
\ee
omitting the sum over states. The last factor stems from the electric
field of frequency $\omega$ describing the first interaction at time $t_2<t_1$.
The decaying exponential obviously describes the decay of interference with
the dephasing rate $\Gamma_{\k_\|l';\k_\|l}$.
The oscillating terms are of the form
$\exp[-i\delta\omega(t_1-t_2)]\,\exp(-i\omega t_1)$ with
$\delta\omega=(E_{\k_\|l'}-E_{\k_\|l})/\hbar\;-\;\omega$.
Thus we expect slow beats
with the detuning frequency $\delta\omega$, which lead to
an initial decay of the signal on a time scale of $(\delta\omega)^{-1}$.
There should be a recurring signal at large delay times, but
this is in practice suppressed by relaxation.
A similar argument can be made for 2PPE using Eq.~(\ref{22.eta4.sh}).
The effect is clearly seen in Fig.~\ref{fig.iP.det} for 2PPE: The width of
the pattern is reduced by the detuning. Its tails also become more
irregular.

Next, we turn to the effect of the band structure. We first discuss
SHG. The SHG yield is determined by a sum over many transitions of different
energies and dephasing rates, see Eq.~(\ref{2.chi210}).
If the decay is governed by dephasing one observes the smallest
dephasing rate at large time delays $\Delta T$. However, at intermediate
$\Delta T$ one sees an averaged rate. The dephasing rate
$\Gamma_{\k_\|l;\k_\|l'}$ for two bands $l$ and $l'$ should usually
not change dramatically with $\k$. On the other hand,
the contribution of detuning is necessarily different for transitions with
different transition frequencies.
Thus, in the interference pattern a continuum of beating frequencies appears.
Consequently, the initial decay is governed by an average detuning and later,
probably unobservable, recurring signals are strongly reduced by destructive
interference of different beating frequencies. The averages are weighted
by a factor approximately inversely proportional to the detuning,
as seen from Eq.~(\ref{21.chiom}). For narrow valence {\it and\/}
intermediate state bands the average is restricted to a small effective
band width $W$.
For broader bands but constant relaxation rates throughout each band the
dependence of numerical results (not shown) for the SHG yield on the width
of the intermediate band turns out to be weak, since in this case all
contributing processes are governed by the same
relaxation rates \cite{rem.SHGband}.
Hence, if only a single intermediate state or a few
states contribute significantly \cite{Luce96}
or if there is a narrow band with uniform relaxation rates, the Bloch
equations should work well. In this case our expressions reduce to a
perturbative solution of the Bloch equations.

On the other hand, if there
is strong electron-electron scattering at certain $\k$ vectors, {\it e.g.},
due to Fermi surface nesting, the rates can be strongly $\k$ dependent.
If many states of different relaxation rates enter the SHG
photon yield, then for broad
bands the description of SHG using optical Bloch equations with a single
intermediate state is not justified. If interference patterns are
fitted with results from Bloch equations, there is no simple relation
between the extracted relaxation rate and the dephasing rates of the excited
electrons.

For 2PPE the situation is quite different, since this method allows to
probe specific momenta $\k$ in the Brillouin zone. Here, the band width is not
crucial. There are typically several contributions to the photoelectron
yield, since there are several unoccupied bands. The contributions are
again weighted with the inverse detuning, but now there is only a small
number of states involved for fixed photoelectron momentum $\k$. Thus a
description in terms of a small number of states, {\it e.g.}, by optical
Bloch equations, is valid. However, if one experimentally integrates
over $\k$, 2PPE behaves much like SFG.

As mentioned above, 2PPE is generally accompanied by SFG, although the
latter is generally smaller in intensity due to the appearance of
additional dipole matrix elements. It might be interesting to perform both
SFG and 2PPE experiments on the same sample. The two techniques are
complementary in that 2PPE gives information about specific points in the
Brillouin zone whereas SFG averages over the whole zone. Furthermore,
comparison of Eqs.~(\ref{2.chi210}) and (\ref{22.eta4.sh}) shows that while
the general form of the expressions for SFG and 2PPE is similar, they do
depend on the material parameters in quite different ways. We give three
examples: First, 2PPE also depends on the energy relaxation rates
(lifetimes) directly, whereas SFG only depends on the dephasing rates.
Secondly, SFG crucially depends on the dipole matrix element ${\bf D}_{31}$
of the transition from the excited state above $\Ev$ to the original state
in the Fermi sea, whereas 2PPE does not. Thus comparison of SFG and 2PPE
may proof useful for measuring the dipole matrix elements. Thirdly, SFG
generally results from a much thinner surface region than 2PPE and the
relaxation rates obtained from 2PPE are more bulk-like, allowing to study
the dependence of the rates on the distance from the surface. Finally, we
have shown that there is a contribution to 2PPE from SFG light generated
within the solid, see Eqs.~(\ref{22.N2red1}) and (\ref{22.N2red2}) as well
as Figs.~\ref{fig.N4}(b) and (c). Simultaneous measurement of SFG and 2PPE
may allow to detect this interesting effect.

\subsection{Collective plasma excitations}
\label{sus.plasmon}

Since SFG and 2PPE may be strongly enhanced by collective plasma
excitations, it is useful to discuss them in the framework of the response
theory. Our goal is to show plasmon enhancement of SFG and 2PPE \emph{in
principle}, even though the plasma frequency is larger than currently
accessible laser frequencies in some metals (but not, for example, in
silver, many heavy-fermion metals, and the interesting compound
$\mathrm{MgB}_2$). Note, the plasma frequency is smaller in clusters,
for which the same general picture applies.

We have seen in Sec.~\ref{sec.theory} that in both SFG and 2PPE
{\it field enhancement\/} of the effective electric field ${\bf E}$ is
described by the Fresnel formulae \cite{SMD87,HBB94,Luce97,LuceLDA,AtK02}.
This mechanism is relevant at the frequency of the exciting laser field
and, in the case of SFG, also at the sum frequency for the outgoing SFG
light. It corresponds to the coupling of the external light field to
\emph{plasmon-polaritons} in the solid \cite{Board}.

The second important origin of plasmon enhancement is the screening of the
nonlinear polarization ${\bf P}^{(2)}$, which is caused by the effective
electric field ${\bf E}^{(2)}$ accompanying the longitudinal part of
$\mathbf{P}^{(2)}$ and appears at the sum frequency. Since the electric
field $\mathbf{E}^{(2)}$ is longitudinal, a true \emph{plasmon} excitation
is involved. It is important to remember that the exciting light does
couple to plasmons in real solids; this coupling is only absent in simple
jellium models \cite{Barton,Board}.

Note, in the case of pump-probe SFG with two pulses of different mean
frequencies $\omega_1$ and $\omega_2$, one of them {\it and\/} the sum
frequency $\omega_1+\omega_2$ can be close to the plasma frequency. This
so-called {\it double resonance\/} leads to a particularly strong
enhancement \cite{Lie99}. {\it Surface\/} plasmons lie outside the scope of
this paper, since they require a more detailed description of the surface.
See Ref.~\cite{Board} for a discussion.

Motivated by 2PPE experiments on clusters \cite{Gerb}, we briefly consider
the plasmon decay. A plasmon decays into a single particle-hole pair
\cite{BVV00}. The probability of this decay is determined by the phase
space available for the final electron-hole pair. It is only energetically
possible if the plasmon dispersion lies within the electron-hole continuum
at the plasmon momentum $\q$, leading to Landau damping. On the other hand,
decay into a single electron-hole pair may be possible close to the
surface, since translational symmetry is broken and $q_z$ is not conserved.
If the energy of the electron is higher than the vacuum energy,
photoemission may result. Creation of several pairs is possible by
subsequent inelastic electron-electron scattering. A plasmon also looses
energy through inelastic scattering of the virtual electrons and holes in
the loop in Fig.~\ref{fig.chi2}(b). This process is governed by the
single-particle relaxation rates. The plasmon lifetime is thus shorter than
typical lifetimes of the relevant excited electrons.

A plasma mode can be multiply excited. In a recent 2PPE experiment a doubly
excited plasma mode of silver nanoparticles decays into a \emph{single}
electron-hole pair \cite{Gerb,rem.clusters}. What is actually observed is
an enhancement of the 2PPE yield when the sum frequency is close to
\emph{twice} the plasma frequency. The origin of this effect is that 2PPE
with the incident-light frequency close to the plasma frequency is enhanced
due to field enhancement. The general process is not specific to clusters
but is also relevant for flat surfaces. It would be interesting to look for
this effect experimentally.

\subsection{Further remarks}


Concerning the range of validity of the second-order response theory we
remark the following. We consider first pump-probe SFG with a very long
delay time $\Delta T$. The first-order density operator $\rho^{(1)}$
describes the result of the first interaction. It contains a finite
polarization (off-diagonal components) but no change of occupation
(diagonal components), see Eq.~(\ref{2.rho1.1}). A change of occupation is
only obtained from $\rho^{(2)}$ and higher-order contributions, which
involve a larger number of dipole matrix elements and are usually small
compared to $\rho^{(1)}$. However, the off-diagonal components usually
decay faster than the diagonal ones so that for long delay times the
higher-order change of occupation can dominate over the second-order
polarization and the second-order approximation breaks down. On the other
hand, our expressions for 2PPE already include the change of occupation due
to the first pulse, since we have directly calculated the photoelectron
yield to fourth order. Thus the results should hold even for long delay
times. For the case that the polarization has decayed at the time of the
second pulse, but the non-equilibrium occupation has not, the resulting
limiting form of $\eta^{\mathrm{2PPE}}$ is given by Eq.~(\ref{ab.e41}). It
only depends on the {\it energy relaxation\/} rates. This is the case where
rate equations are appropriate \cite{Kno00,Rol1,Rol2}. Due to the vanishing
polarization there are no interference effects.

There is an alternative and physically appealing description of pump-probe
SFG and 2PPE as a {\it two-step process\/}: The first pulse creates a
non-equilibrium distribution, which is probed by the second one. We now
discuss the validity of calculations based on this picture. In
App.~\ref{app.a} we derive an expression for the linear susceptibility of
an electron gas in an arbitrary non-equilibrium state described by the
density matrix $\rho_{\mathrm{neq}}$, see Eq.~(\ref{2.chi1.7}). If we
insert $\rho^{(1)}$ due to the first pulse for $\rho_{\mathrm{neq}}$, we
obtain a two-step description for $\chi^{(2)}$ and the polarization ${\bf
P}^{(2)}$. Omitting the details, we only state that the result is identical
to the one obtained directly for the second-order polarization ${\bf
P}^{(2)}$, Eq.~(\ref{21.P2}), but with the full susceptibility $\chi^{(2)}$
replaced by its irreducible part $\chi^{(2)}_{\mathrm{irr}}$ of
Eq.~(\ref{2.chi210}). Thus, by assuming two separate interaction processes
and treating each in a first-order approximation, we loose the screening of
the second-order polarization. This is not justified if the sum frequency
lies close to the plasma resonance.

Next we consider a two-step description of 2PPE: 
The 2PPE photoelectron yield ${\cal N}^{\mathrm{2PPE}}$ is expressed in terms of an
arbitrary non-equilibrium density matrix $\rho_{\mathrm{neq}}$
as discussed in App.~\ref{app.b}. Then the second-order density
matrix $\rho^{(2)}$ due to the first pulse is inserted for
$\rho_{\mathrm{neq}}$. We reobtain the
full irreducible fourth-order result ${\cal N}^{\mathrm{2PPE}}_{\mathrm{irr}}$
of Fig.~\ref{fig.N4}(a), but only part of the reducible contributions,
Fig.~\ref{fig.N4}(b), (c): The two-step description
neglects contributions of two photons out of {\it different\/} pulses
being converted into one SHG photon. These contributions
may become important if the sum frequency is close to a plasma resonance.
In conclusion, the two-step picture of SFG and 2PPE is valid unless the
response at the sum frequency is enhanced by plasmon effects.

Finally, we emphasize that our theory can also describe time-resolved SFG
and 2PPE from ferromagnetically ordered systems. Ultimately, the light
couples to the (spin) magnetization through spin-orbit coupling, which is
incorporated, in principle, in the dipole matrix elements ${\bf D}$ and the
band structure. The spin-dependent matrix elements can be calculated by a
perturbative expansion in the spin-orbit coupling \cite{HB,NOLI}. SFG and
2PPE also depend on magnetic order through the band energies $E_{\k_\| l}$
and relaxation rates $\Gamma_{\k_\|l;\k_\|'l'}$, since $l$ also contains a
spin index $\sigma$. Of particular importance for magnetically-ordered
materials is the rotation of the polarization of SHG light relative to
incident light (NOLIMOKE) \cite{HB,NOLI,Luce96}. As mentioned above, the
light polarization is controlled by the symmetries of the tensor
$\chi^{(2)}_{ijk}$, which are known for low-index surfaces \cite{Pan}.

Compared to NOLIMOKE, 2PPE for magnetic systems has the advantage that in
principle one can obtain information on the spin-dependent lifetimes of
electrons in specific states $|\k_\|l\rangle$. The dependence of 2PPE on
the light polarization for magnetic systems has not been studied so far. In
the response theory this dependence is controlled by the symmetries of the
tensor $\eta^{\mathrm{2PPE}}_{ijkl}$, as mentioned in Sec.~\ref{ss.22}.

For pump-probe experiments with \emph{long} time delay $\Delta T$ the main
contribution to 2PPE comes from the change of occupation brought about by
the pump pulse. \emph{Only} in this case the photoelectron yield is
proportional to the occupation of the corresponding intermediate states. If
in addition the matrix elements and the relaxation rates out of vacuum
states depend only weakly on spin, then Eq.~(\ref{4.Nrho.1}) shows that the
2PPE yield becomes proportional to the spin-dependent occupation of these
intermediate states:
\be
{\cal N}^{\mathrm{2PPE}}(\Delta T;\k,\sigma) \propto
  \rho_{\mathrm{neq};\k\nu\sigma;\k\nu\sigma}(\Delta T) =
  n_{\mathrm{neq};\k\nu\sigma}(\Delta T) ,
\ee
where $n_{\mathrm{neq};\k\nu\sigma}$ denotes the non-equilibrium occupation
of the state $|\k\nu\sigma\rangle$ after the pump pulse. Then the
difference of the spin-up and spin-down 2PPE yield, ${\cal
N}^{\mathrm{2PPE}}(\Delta T;\k,\uparrow)-{\cal N}^{\mathrm{2PPE}}(\Delta
T;\k,\downarrow)$, is proportional to the difference of the occupations and
thus to the transient magnetization of the intermediate states.

Note, circularly polarized light might excite electrons spin-selectively
due to angular-momentum conservation. In our response theory these
selection rules are incorporated in the dipole matrix elements
$\mathbf{D}$. Conversely, spin-selective excitation will lead to
corresponding polarization of the SFG light. The use of circularly
polarized light in 2PPE and SFG is of particular interest regarding
ferromagnets and transient magnetizations.

\subsection{Conclusions}


To summarize, we have presented a unified perturbative response theory for
time-resolved SFG and 2PPE. The theory is fully quantum-mechanical and
contains the interference effects described by off-diagonal components of
the density matrix. It does not rely on any assumption about the time or
frequency dependence of the exciting laser pulses. The solid is described
by its band structure, relaxation rates, and dipole matrix elements. We
have discussed metals but the response theory can be applied to
semiconductors and insulators as well, see, \textit{e.g.},
Ref.~\cite{ShA03}. Since the theory is formulated directly in the time
domain, it presents a suitable framework for the discussion of the
time-dependent physics of SFG and 2PPE. We have shown that similar
information as from 2PPE can be gained from SHG. Of course, 2PPE is
sensitive to specific momenta $\k$ in the Brillouin zone, while SHG in
general is not. A simple tight-binding model of a metal has been studied in
order to show that the theory gives reasonable numerical results and to
illustrate the following effects important for the understanding of SFG and
2PPE.

We have shown how relaxation rates and detuning affect the interference
patterns in single-color pump-probe SHG and 2PPE experiments: The lifetime
in the intermediate states and their detuning with respect to the photon
energy lead to a similar narrowing of the interference patterns. The effect
of detuning must be taken into account in order to extract meaningful life
times from such experiments. Also, in particular in SHG the measured
relaxation rate is a weighted average over the relaxation rates of many
excited states. Furthermore, the weights in this average change with the
pump-probe delay. Thus different rates govern the decay of the interference
pattern depending on the pump-probe delay---the decay is not simply
exponential. We have also discussed the range of validity of the optical
Bloch equations, applicable if only a few states contribute, and of
semiclassical rate equations valid for very long pump-probe delays. Both
approaches are limiting cases of our theory.

Finally, we have considered the role played by collective plasma
excitations. Plasmon effects in both SFG and 2PPE can only partly be
understood in terms of field enhancement at the surface. One also has to
take the electric field accompanying a nonlinear polarization of the
electron system into account. This effect leads to interesting additional
contributions to 2PPE, in which incoming photons are converted into
sum-frequency photons which then lead to \emph{ordinary} photoemission.
These contributions should be observable if the sum frequency is close to
the plasma frequency.

\acknowledgments

We thank Drs.\ W. Pfeiffer, G. Gerber, G. Bouzerar, and R. Knorren for
valuable discussions. Financial support by the Deutsche
Forschungsgemeinschaft through Sonderforschungsbereich 290 is gratefully
acknowledged.

\appendix

\section{Response theory for the nonlinear optical response}
\label{app.a}

In this appendix we derive the transverse second-order susceptibility and
polarization for arbitrary pulse shapes of the exciting laser field. The
resulting expressions allow to calculate the SFG
yield for arbitrary pulse shapes, thereby going beyond the results of
H\"ubner and Bennemann \cite{HB} for continuous-wave, monochromatic light.
The self-consistent-field approach \cite{EC} is applied to a solid described
by its band structure and relaxation rates. The flat surface is assumed
to lie at $z=0$ with the solid at $z<0$. We neglect the intraband
contribution, which is reasonable for optical frequencies.

The single-particle Hamiltonian is $H = H_0 + V$, where $H_0$ describes the
unperturbed solid with the normalized eigenstates $|\k_\| l\rangle$ and
eigenenergies $E_{\k_\| l}$. $\k_\|$ is the crystal momentum parallel to the
surface and all other quantum numbers, discrete as well as continuous, are
collectively denoted by $l$ (see the discussion at the beginning of
Sec.~\ref{ss.21}).

The time-dependent perturbation is \cite{Schiff}
\be
V(\r,t) = -\frac{ie\hbar}{mc}\, {\bf A}(\r,t)\cdot\Nabla
 - \frac{ie\hbar}{2mc}\, [\Nabla\cdot{\bf A}(\r,t)]
\ee
with the vector potential ${\bf A}$, which is treated classically.
We have made the usual approximation to
neglect the quadratic term in ${\bf A}$ and have used a gauge with
vanishing scalar potential. We will later need the temporal Fourier
transform (using the convention of Ref.~\cite{HB})
\ba
V(\r,\omega) & = & \int \frac{dt}{2\pi}\, e^{-i\omega t}\, V(\r,t)
  \nonumber \\
& = & \frac{e\hbar}{m\omega}\, {\bf E}(\r,\omega)\cdot\Nabla
  + \frac{e\hbar}{2m\omega}\, [\Nabla\cdot{\bf E}(\r,\omega)] ,
\ea
where we have used ${\bf E}=-(1/c)\,\partial {\bf A}/\partial t$. Inserting
the spatial Fourier transform we get
\be
V(\r,\omega) = \frac{e\hbar}{m\omega} \sum_\q e^{-i\q\cdot\r}\,
  {\bf E}(\q,\omega) \cdot \left(\Nabla - \frac{i\q}{2}\right) .
\ee
The matrix elements of $V$ are
\ba
\langle \k_\| l|V|\k_\|+\q_\|, l'\rangle
& = & \frac{e\hbar}{m\omega} \sum_{q_z} {\bf E}(\q,\omega)\cdot
  \langle \k_\| l| e^{-i\q\cdot\r} \left(\Nabla-i\q/2\right)
  |\k_\|+\q_\|, l'\rangle \nonumber \\
& \equiv & e \sum_{q_z} {\bf E}(\q,\omega)\cdot
  {\bf D}_{\k_\| l;\k_\|+\q_\|,l'}(q_z,\omega)
\label{1.VD.1}
\ea
using momentum conservation and $\q=(\q_\|,q_z)$. Here,
\be
{\bf D}_{\k_\| l;\k_\|+\q_\|,l'}(q_z,\omega)
= \frac{\hbar}{m\omega}\,
  \langle \k_\| l| e^{-i\q\cdot\r} \left(\Nabla-i\q/2\right)
  |\k_\|+\q_\|, l'\rangle .
\ee
If the field $\mathbf{E}$
were purely transverse the term $i\q/2$ in the parentheses
would drop out, but this is not guaranteed close to a surface. It is not our
goal to calculate ${\bf D}$ explicitly. We only remark
that if one uses the dipole approximation
\cite{HK90,HBB94,Luce97,LuceLDA,NOLI,Luce96,Luce98,AnH02,Shala}
$e^{-i\q\cdot\r}\cong 1$, the contribution from $i\q/2$ vanishes since we
neglect the intraband contributions so that $l'\neq l$,
and the remainder gives \cite{Klings,Voon}
\ba
\lefteqn{
\frac{\hbar}{m\omega}\, \langle \k_\| l|
  \Nabla |\k_\|+\q_\|, l'\rangle = -\frac{1}{\hbar\omega}\, \langle \k_\| l|
  [H, \r] |\k_\|+\q_\|, l'\rangle } \nonumber \\
& & = -\frac{E_{\k_\| l} - E_{\k_\|+\q_\|,l'}}{\hbar\omega}\,
  \langle \k_\| l| \r |\k_\|+\q_\|, l'\rangle .
\label{1.Ddipol.2}
\ea
Since the response is dominated by contributions with $\hbar\omega\sim
E_{\k_\|+\q_\|,l'}-E_{\k_\| l}$ the prefactor can be further approximated
by unity if the frequency spectrum of the incoming light is sufficiently
narrow. The dipole approximation should be justified since the electric
field changes slowly on the scale of the lattice constant (the skin depth
is about one order of magnitude larger than the lattice constant).
However, our
response theory for SFG does not require the dipole approximation to be
made.

The time evolution of the density operator $\rho$ is described by the
master or von Neumann equation \cite{Louisell,Loudon}
\be
\frac{d}{dt}\rho = \frac1{i\hbar}\,[H,\rho] + {\cal R}[\rho] .
\label{aa.master1}
\ee
The functional ${\cal R}[\rho]$ represents
relaxation terms made explicit below. Matrix elements of $\rho$ are
written as
$\rho_{\k_\| l;\k'_\|l'} \equiv \langle \k_\| l| \rho |\k'_\| l'\rangle$.
The master equation then reads \cite{Loudon}
\ba
\lefteqn{
\frac{d}{dt} \rho_{\k_\| l;\k'_\|l'}
  = \frac1{i\hbar} \langle \k_\|l| [H_0+V,\rho] |\k'_\|l'\rangle
  + \delta_{\k_\|\k'_\|} \delta_{ll'}
  } \nonumber \\
& & \quad\times
  {\sum_{\k''_\|l''}}^\prime
      \gamma_{\k_\|l;\k''_\|l''} \rho_{\k''_\|l'';\k''_\|l''}
    - \Gamma_{\k_\|l;\k'_\|l'} \rho_{\k_\|l;\k'_\|l'} .
\label{1.vN}
\ea
Here, $\Gamma_{\k_\|l;\k_\|l} \equiv \tau_{\k_\|l}^{-1}$ is the inverse
lifetime of state $|\k_\|l\rangle$, which arises mainly from
inelastic electron-electron scattering.
$\gamma_{\k_\| l;\k'_\|l'}$ gives the rate of spontaneous transitions from
state $|\k'_\|l'\rangle$ to state $|\k_\| l\rangle$. Because of conservation
of electron number
\be
\Gamma_{\k_\|l;\k_\|l} = {\sum_{\k'_\|l'}}^\prime \gamma_{\k'_\|l';\k_\|l} .
\label{1.gammasum}
\ee
Primed sums run over all states except $|\k_\|l\rangle$.
$\Gamma_{\k_\|l;\k_\|l}$ and $\gamma_{\k_\| l;\k'_\|l'}$ describe
{\it energy relaxation}, {\it i.e.}, the change of the diagonal components
of $\rho$, whereas the {\it dephasing\/} rate $\Gamma_{\k_\| l;\k'_\|l'}$
with $|\k_\| l\rangle \neq |\k'_\|l'\rangle$
describes relaxation of the off-diagonal components.

To solve the master equation (\ref{1.vN}) perturbatively, the
density operator is expanded in powers of the perturbation $V$ as
$\rho=\rho^{(0)}+\rho^{(1)}+\rho^{(2)}+\ldots$ In thermal equilibrium
the unperturbed density matrix is expressed in terms of the
Fermi function,
$\rho^{(0)}_{\k_\| l;\k_\|'l'} = \delta_{\k_\|\k_\|'} \delta_{ll'}
  f(E_{\k_\| l})$.
The temporal Fourier transform of Eq.~(\ref{1.vN}) reads
\ba
\lefteqn{
i\omega \rho_{\k_\| l;\k'_\|l'}(\omega)
  = \left( \frac{E_{\k_\| l} - E_{\k'_\| l'}}{i\hbar}
  - \Gamma_{\k_\|l;\k'_\|l'} \right)\,  \rho_{\k_\| l;\k'_\|l'}(\omega) }
  \nonumber \\
& & {}+ \frac{1}{i\hbar} \int \frac{dt}{2\pi}\: e^{-i\omega t}
  \sum_{\k''_\| l''} \left[ \langle \k_\|l|V(t)|\k''_\|l''\rangle\,
  \rho_{\k''_\|l'';\k'_\|l'}(t) - \rho_{\k_\|l;\k''_\|l''}(t)\,
  \langle \k''_\|l''|V(t)|\k'_\|l'\rangle \right]
  + \delta_{\k_\|\k'_\|} \delta_{ll'} {\sum_{\k''_\|l''}}^\prime
  \gamma_{\k_\|l;\k''_\|l''} \rho_{\k''_\|l'';\k''_\|l''}(\omega) .
\label{1.vN.F1}
\ea
Keeping only terms linear in $V$ one obtains
\ba
\lefteqn{
\rho^{(1)}_{\k_\| l;\k_\|+\q_\|,l'}(\omega)
  = e\, \sum_{q_z}
    {\bf D}_{\k_\| l;\k_\|+\q_\|,l'}(q_z,\omega)
    \cdot {\bf E}(\q,\omega)
  } \nonumber \\
& & \quad\times
  \frac{f(E_{\k_\|+\q_\|,l'}) - f(E_{\k_\| l})}
    {-\hbar\omega + E_{\k_\|+\q_\|,l'} - E_{\k_\| l}
      + i\hbar\Gamma_{\k_\| l;\k_\|+\q_\|,l'}} .
\label{2.rho1.1}
\ea
Note that the diagonal components vanish: There is no change of occupation to
first order.

The polarization is given by the thermal average of $-e{\bf D}$, which is the
conjugate of the electric field according to Eq.~(\ref{1.VD.1}),
\ba
{\bf P}(\q,\omega) & = & -\frac{e}{v} \sum_{\k_\|} \sum_{ll'}
  \rho_{\k_\| l;\k_\|+\q_\|,l'}(\omega)
  \: {\bf D}_{\k_\|+\q_\|,l';\k_\| l}(-q_z,\omega) ,\!\!
\label{2.P.1}
\ea
where ${v}$ is the volume. To first order
\be
P_i^{(1)}(\q_\|,q_z,\omega) = \sum_{q'_z}
  \chi_{ij}^{(1)}(\q_\|,q_z,q'_z,\omega)\,
  E_j(\q_\|,q'_z,\omega) ,
\label{2a.P1.2}
\ee
with the linear susceptibility of Lindhard form
\ba
\lefteqn{
\chi_{ij}^{(1)}(\q_\|,q_z,q'_z,\omega) = -\frac{e^2}{v} \sum_{\k_\|}
  \sum_{ll'} D^i_{\k_\|+\q_\|,l';\k_\|l}(-q_z,\omega)
  } \nonumber \\
& & \quad {}\times
  D^j_{\k_\|l;\k_\|+\q_\|,l'}(q'_z,\omega)
  \nonumber \\
& & \quad {}\times
  \frac{f(E_{\k_\|+\q_\|,l'}) - f(E_{\k_\| l})}
    {-\hbar\omega + E_{\k_\|+\q_\|,l'} - E_{\k_\| l}
      + i\hbar\Gamma_{\k_\| l;\k_\|+\q_\|,l'}} ,
\label{2.chi1.1}
\ea
shown diagrammatically in Fig.~\ref{fig.chi1}. It takes into account that
the $z$ component of momentum is not conserved.

If we neglect the frequency dependence of $\mathbf{D}$
we obtain a result in the time domain. Equation (\ref{2a.P1.2}) and the
Fourier transform of Eq.~(\ref{2.chi1.1}) give
\ba
P_i^{(1)}(\q_\|,q_z,t) & = & \sum_{q'_z}
  \int_{-\infty}^\infty \frac{dt_1}{2\pi}\,
  \chi_{ij}(\q_\|,q_z,q'_z,t-t_1)
  \nonumber \\
& & \times E_j(\q_\|,q'_z,t_1) ,
\ea
with
\ba
\lefteqn{
\chi_{ij}(\q_\|,q_z,q'_z;t-t_1)
  = \frac{e^2}{v}\,\frac{2\pi i}{\hbar}\,\Theta(t-t_1)
  \sum_{\k_\|} \sum_{ll'} D^i_{\k_\|+\q_\|,l';\k_\| l}(-q_z)\,
  D^j_{\k_\| l;\k_\|+\q_\|,l'}(q'_z)
  } \nonumber \\
& & \quad {}\times 
  [f(E_{\k_\|+\q_\|,l'})-f(E_{\k_\| l})]\:
  \exp\!\left[i \frac{E_{\k_\|+\q_\|,l'}-E_{\k_\| l}}
    {\hbar}(t-t_1)\right]\:
  \!\exp[-\Gamma_{\k_\| l;\k_\|+\q_\|,l'}(t-t_1)] .
\ea
The response is thus only non-zero if $t>t_1$, which expresses
causality.

The above results have been obtained under the assumption that the
system is initially in thermal equilibrium.
We now consider the non-equilibrium case.
Then the unperturbed polarization ${\bf P}_{\mathrm{neq}}^{(0)}$ is, in
general, non-vanishing so that the electrons experience an effective
field even in the absence of an external perturbation,
leading to a master equation that is nonlinear in the
non-equilibrium density operator $\rho_{\mathrm{neq}}^{(0)}$.
We assume, however, that this effect of electron-electron interaction
is negligible. Then the linear response of a non-equilibrium system can be
written as
\be
P_{\mathrm{neq};i}^{(1)}(\q_\|,q_z,\omega) = \!\sum_{\q'_\|,q'_z}
  \int_{-\infty}^\infty \!\!\! d\omega'\,
  \chi_{\mathrm{neq};ij}^{(1)}(\q_\|,q_z,\omega;\q'_\|,q'_z,\omega')\,
  E_j(\q',\omega') ,
\label{2.P1gen1}
\ee
with
\ba
\lefteqn{
\chi_{\mathrm{neq};ij}^{(1)}(\q_\|,q_z,\omega;\q'_\|,q'_z,\omega')
  = -\frac{e^2}{v} \sum_{\k_\|} \sum_{ll'l''}
  D^i_{\k_\|+\q_\|,l';\k_\|l}(-q_z,\omega) } \nonumber \\
& & \quad {}\times
  \Big[ D^j_{\k_\|l;\k_\|+\q'_\|,l''}(q'_z,\omega')\,
   \rho^{(0)}_{\mathrm{neq};\k_\|+\q'_\|,l'';\k_\|+\q_\|,l'}(\omega-\omega')
  \nonumber \\
& & \qquad {}- \rho^{(0)}_{\mathrm{neq};\k_\|l;\k_\|+\q_\|-\q'_\|,l''}
      (\omega-\omega')\,
    D^j_{\k_\|+\q_\|-\q'_\|,l'';\k_\|+\q_\|,l'}(q'_z,\omega') \Big] .
\label{2.chi1.7}
\ea
This equation gives the linear susceptibility in terms of an
{\it arbitrary\/} density operator $\rho_{\mathrm{neq}}^{(0)}$.

To return to the response of an equilibrium system, we now consider the
second-order contribution.
We collect the terms in the master equation (\ref{1.vN.F1})
that are of second order in the effective electric field,
\ba
i\omega \rho^{(2)}_{\k_\|l;\k_\|+\q_\|,l'}
  & = & \frac1{i\hbar}
  (E_{\k_\|l}-E_{\k_\|+\q_\|,l'}) \rho^{(2)}_{\k_\|l;\k_\|+\q_\|,l'}
  \nonumber \\
& & {}+ \frac1{i\hbar} \langle\k_\|l| [V^{(1)},\rho^{(1)}]
    |\k_\|+\q_\|,l'\rangle
  + \frac1{i\hbar} \langle\k_\|l| [V^{(2)},\rho^{(0)}] |\k_\|+\q_\|,l'\rangle
  \nonumber \\
& & {}+ \delta_{\q_\| 0} \delta_{ll'} {\sum_{\kap_\|\lambda}}^\prime
    \gamma_{\k_\|l;\kap_\|\lambda} \rho^{(2)}_{\kap_\|\lambda;\kap_\|\lambda}
  - \Gamma_{\k_\|l;\k_\|+\q_\|,l'} \rho^{(2)}_{\k_\|l;\k_\|+\q_\|,l'} .
\label{3.vN1}
\ea
$V^{(1)}\equiv V$ is the perturbation by the effective field.
Higher-order perturbations $V^{(n)}$, $n\ge 2$, result from the electric
field due to the displaced charge calculated at order $n$. From the
expression for the electric field due to a polarization
${\bf P}^{(n)}$, Eq.~(\ref{21.Jackson2}) \cite{Jackson}, one obtains
${\bf E}^{(n)}(\q) = -4\pi\,\hat\q\,\hat\q\cdot{\bf P}^{(n)}(\q)$,
where $\hat\q\equiv \q/|\q|$. Importantly, this additional field also has to
be taken into account as a perturbation \cite{HB}. Specifically,
the longitudinal part of
${\bf P}^{(n)}$ leads to a perturbation with matrix elements
\be
\langle \k_\| l|V^{(n)}|\k_\|+\q_\|, l'\rangle
  = e \sum_{q_z} {\bf E}^{(n)}(\q,\omega)\cdot
  {\bf D}_{\k_\| l;\k_\|+\q_\|,l'}(q_z,\omega) ,
\label{aa.Vn1}
\ee
see Eq.~(\ref{1.VD.1}). Note, due to the reduced symmetry at the
surface a transverse electric field in general leads to a polarization
with a longitudinal component.

Since the field $\mathbf{E}^{(2)}=-4\pi\,\hat\q\,\hat\q\cdot{\bf P}^{(2)}$
is explicitly of \emph{second} order in the applied field, see
Eq.~(\ref{21.P2}), it must be taken into account in our calculation of the
second-order response.
From Eqs.~(\ref{2.P.1}) and (\ref{3.vN1}) we then obtain for $\q_\|\neq 0$
\ba
P^{(2)}_i(\q,\omega) & = & -\frac{e^2}{v}\, \int d^2k_\| \sum_{ll'}
  \frac{D^i_{\k_\|+\q_\|,l';\k_\| l}(-q_z,\omega)}
    {-\hbar\beta+E_{\k_\|+\q_\|,l'}
    -E_{\k_\| l} + i\Gamma_{\k_\| l;\k_\|+\q_\|,l'}} \nonumber \\
& & \quad{}\times \int d^3q'\, \sum_{\lambda} \int d\omega'
  \bigg[ D^j_{\k_\| l;\k_\|+\q'_\|,\lambda}(q'_z,\omega')\,
    \rho^{(1)}_{\k_\|+\q'_\|,\lambda;\k_\|+\q_\|,l'}(\omega-\omega')
    \nonumber \\
& & \qquad{}- \rho^{(1)}_{\k_\| l;\k_\|+\q_\|-\q'_\|,\lambda}(\omega-\omega')\,
    D^j_{\k_\|+\q_\|-\q'_\|,\lambda;\k_\|+\q_\|,l'}(q'_z,\omega') \bigg]\,
  E_j(\q',\omega')  \nonumber \\
& & {}+ 4\pi\left[f(E_{\k_\|+\q_\|,l'}) - f(E_{\k_\| l})\right]
  \int dq_z'\, D^j_{\k_\| l;\k_\|+\q_\|,l'}(q'_z,\omega)\,\hat{q}_j
  \hat{q}_k P^{(2)}_k(\q',\omega) ,
\ea
where in the last term $\q'_\|=\q_\|$. Note that the nonlinear polarization
$\mathbf{P}^{(2)}$ appears on both sides of the equation. Solving this
equation for ${\bf P}^{(2)}$ we obtain
\be
P^{(2)}_i(\q,\omega) = \sum_{\q',\q''}
  \int_{-\infty}^\infty \!\!\! d\omega'\,
  \chi^{(2)}_{ijk}(\q,\q';\omega,\omega')\,
  E_j(\q',\omega')\, E_k(\q-\q',\omega-\omega') ,
\label{2.P24}
\ee
with the second-order susceptibility
\be
\chi^{(2)}_{ijk}(\q,\q';\omega,\omega')
 = \sum_m \sum_{\kappa_z}\,
  \varepsilon^{-1}_{\mathrm{long};im}(\q_\|,q_z,\kappa_z,\omega) \:
  \chi^{(2)}_{\mathrm{irr};mjk}\big((\q_\|,\kappa_z),
    \q',\q-\q';\omega,\omega'\big)
\label{3.chiom}
\ee
and
\be
\varepsilon_{\mathrm{long};ij}(\q_\|,q_z,\kappa_z,\omega)
  \equiv \delta_{ij} + 4\pi \chi_{im}(\q_\|,q_z,\kappa_z,\omega)\,
  \frac{(\q_\|,\kappa_z)_k}{|(\q_\|,\kappa_z)|}\,
  \frac{(\q_\|,\kappa_z)_j}{|(\q_\|,\kappa_z)|} .
\label{3.epslong1}
\ee
Here, $\varepsilon_{\mathrm{long};ij}^{-1}(\q_\|,q_z,\kappa_z,\omega)$ is
the \emph{inverse matrix} of $\varepsilon_{\mathrm{long}}$ with respect to
the indices $(i,q_z)$ and $(j,\kappa_z)$. Equation (\ref{3.epslong1}) means
that $\varepsilon_{\mathrm{long}}$ only acts on the longitudinal component
and is unity for the transverse ones. Solving for $\mathbf{P}^{(2)}$ is
found to be equivalent to the summation of an RPA series for the
electron-electron interaction mediated by the electromagnetic field. In this
language the interaction is absorbed into $\chi_{im}$ in
Eq.~(\ref{3.epslong1}).

The irreducible susceptibility in Eq.~(\ref{3.chiom}) reads
\ba
\lefteqn{
\chi^{(2)}_{\mathrm{irr};ijk}(\q,\q',\q'';\omega,\omega')
 = -\frac{e^3}{v}\, \sum_{\k_\|} \sum_{ll'l''}
  \frac{D^i_{\k_\|+\q_\|,l;\k_\| l''}(-q_z,\omega)}{-\hbar\omega
  +E_{\k_\|+\q_\|,l}-E_{\k_\| l''}+i\hbar\Gamma_{\k_\| l'';\k_\|+\q_\|,l}}
  } \nonumber \\
& & \; {}\times \bigg[
  D^j_{\k_\| l'';\k_\|+\q_\|',l'}(q'_z,\omega')\,
  D^k_{\k_\|+\q_\|',l';\k_\|+\q_\|,l}(q''_z,\omega-\omega')\,
  \frac{f(E_{\k_\|+\q_\|,l})-f(E_{\k_\|+\q_\|',l'})}
    {-\hbar\omega+\hbar\omega'+E_{\k_\|+\q_\|,l}-E_{\k_\|+\q_\|',l'}
      +i\hbar\Gamma_{\k_\|+\q_\|',l';\k_\|+\q_\|,l}}
  \nonumber \\
& & \quad {}- D^j_{\k_\|+\q_\|-\q_\|',l';\k_\|+\q_\|,l}(q'_z,\omega')\,
  D^k_{\k_\| l'';\k_\|+\q_\|-\q_\|',l'}(q''_z,\omega-\omega')\,
  \frac{f(E_{\k_\|+\q_\|-\q_\|',l'})-f(E_{\k_\| l''})}
    {-\hbar\omega+\hbar\omega'+E_{\k_\|+\q_\|-\q_\|',l'}-E_{\k_\| l''}
      +i\hbar\Gamma_{\k_\| l'';\k_\|+\q_\|-\q_\|',l'}}
  \bigg] ,
\label{99.chi26}
\ea
shown diagrammatically in Fig.~\ref{fig.chi2}(a).
Finally, the time dependence of the polarization ${\bf P}^{(2)}$
is obtained by Fourier transformation of Eq.~(\ref{2.P24}) using
Eq.~(\ref{2.chi21}), leading to Eq.~(\ref{2.chi210}).

\section{Response theory for photoemission}
\label{app.b}

In this appendix we give details of the derivation of the
time-integrated photoelectron yield.
We also present the analytical expressions for the response
functions omitted in Sec.~\ref{ss.22}.
The starting point is again the master equation
(\ref{aa.master1}). The terms of order $n\ge 1$
can be calculated recursively,
\ba
\lefteqn{
\frac{d}{dt} \rho^{(n)}_{\k_\| l;\k_\|+\q_\|,l'} = \frac1{i\hbar}
  (E_{\k_\| l}-E_{\k_\|+\q_\|,l'}) \rho^{(n)}_{\k_\| l;\k_\|+\q_\|,l'}
  } \nonumber \\
& & \quad {}+ \frac1{i\hbar} \sum_{m=1}^n
  \langle\k_\| l| [V^{(m)},\rho^{(n-m)}] |\k_\|+\q_\|,l'\rangle
  \nonumber \\
& & \quad {}+ \delta_{\q_\| 0} \delta_{ll'} {\sum_{\kap_\|\lambda}}^\prime
    \gamma_{\k_\| l;\kap_\|\lambda} \rho^{(n)}_{\kap_\|\lambda;\kap_\|\lambda}
  \nonumber \\
& & \quad {}- \Gamma_{\k_\| l;\k_\|+\q_\|,l'}
  \rho^{(n)}_{\k_\| l;\k_\|+\q_\|,l'} .
  \qquad\qquad\qquad
\label{4.mastern.1}
\ea
Here, $V^{(m)}$ is the perturbation of order $m$, see the discussion leading
to Eq.~(\ref{aa.Vn1}). Among the states $|\k_\| l\rangle$
etc.\ appearing in Eq.~(\ref{4.mastern.1}) are
states $|\k\sigma,\iv\rangle$ lying in the crystal above the vacuum level.
We assume that electrons leaving the crystal are in states
$|\k\sigma,\ou\rangle$ and are detected without further
interaction and without returning to the solid. Then the
only way their occupation can change
is through spontaneous transitions out of $|\k\sigma,\iv\rangle$,
governed by the rate $\gamma_{\k\sigma,\ou;\k\sigma,\iv}$.
Note, in principle higher vacuum bands appear by shifting the (nearly)
free electron dispersion back into the first Brillouin zone. We omit
these bands for notational simplicity.

First, we consider the irreducible part $\rho^{(n)}_{\mathrm{irr}}$.
This is the contribution of only the direct, first-order perturbation
$V^{(1)}=V$ at every step of the recursion. The resulting equation reads
\ba
\lefteqn{
\frac{d}{dt} \rho^{(n)}_{\mathrm{irr};\k_\| l;\k_\|+\q_\|,l'}(t) = \frac1{i\hbar}
  (E_{\k_\| l}-E_{\k_\|+\q_\|,l'})\,
  \rho^{(n)}_{\mathrm{irr};\k_\| l;\k_\|+\q_\|,l'}(t)
  } \nonumber \\
& & \quad {}+ \frac{e}{i\hbar}
  \sum_{\q'\lambda} \Big[ {\bf D}_{\k_\| l;\k_\|+\q'_\|,\lambda}(q'_z)\,
    \rho^{(n-1)}_{\mathrm{irr};\k_\|+\q'_\|,\lambda;\k_\|+\q_\|,l'}(t)
  - \rho^{(n-1)}_{\mathrm{irr};\k_\| l;\k_\|+\q_\|-\q'_\|,\lambda}(t)\,
    {\bf D}_{\k_\|+\q_\|-\q'_\|,\lambda;\k_\|+\q_\|,l'}(q'_z) \Big]
  \cdot {\bf E}(\q',t)
  \nonumber \\
& & \quad {}+ \delta_{\q_\| 0} \delta_{ll'} {\sum_{\kap_\|\lambda}}^\prime
    \gamma_{\k_\| l;\kap_\|\lambda}\,
    \rho^{(n)}_{\mathrm{irr};\kap_\|\lambda;\kap_\|\lambda}(t)
  - \Gamma_{\k_\| l;\k_\|+\q_\|,l'}\,
    \rho^{(n)}_{\mathrm{irr};\k_\| l;\k_\|+\q_\|,l'}(t) .
\label{4.mastern.2}
\ea
We assume $\mathbf{D}$ to be frequency-independent, see App.~\ref{app.a}.
In Eq.~(\ref{4.mastern.2})
we write the first sum in the form $\sum_{\q'\lambda}$ as a reminder that
{\it all\/} components of the external momentum $\q'$ are summed over.
Hence, we here exclude $q'_z$ from $\lambda$.
The solution for the off-diagonal elements is
\ba
\lefteqn{
\rho^{(n)}_{\mathrm{irr};\k_\| l;\k_\|+\q_\|,l'}(t) = \frac{e}{i\hbar}
  \int_{-\infty}^t \!\!\! dt_1\, \sum_{\q'\lambda}
  e^{-\Gamma_{\k_\| l;\k_\|+\q_\|,l'}(t-t_1)}\,
  \exp\!\left[-i\frac{E_{\k_\| l}-E_{\k_\|+\q_\|,l'}}{\hbar}(t-t_1)\right]
  } \nonumber \\
& & \quad {}\times
  \Big[
  {\bf D}_{\k_\| l;\k_\|+\q'_\|,\lambda}(q'_z)\,
  \rho^{(n-1)}_{\mathrm{irr};\k_\|+\q'_\|,\lambda;\k_\|+\q_\|,l'}(t_1)
  - \rho^{(n-1)}_{\mathrm{irr};\k_\| l;\k_\|+\q_\|-\q'_\|,\lambda}(t_1)\,
  {\bf D}_{\k_\|+\q_\|-\q'_\|,\lambda;\k_\|+\q_\|,l'}(q'_z) \Big]
  \cdot {\bf E}(\q',t_1) .
\label{4.rhon.odi1}
\ea
For the diagonal components there is an additional contribution from the
relaxation of secondary electrons into the given state out of higher-energy
states \cite{Rol1,Rol2}, {\it i.e.},
the term with $\q_\|=0$, $l'=l$ in Eq.~(\ref{4.mastern.2}).
The diagonal components can be written in the implicit form
\ba
\lefteqn{
\rho^{(n)}_{\mathrm{irr};\k_\| l;\k_\| l}(t) = \frac{e}{i\hbar}
  \int_{-\infty}^t \!\!\! dt_1\, \sum_{\q'\lambda}
  e^{-\Gamma_{\k_\| l;\k_\| l}(t-t_1)}\,
  \Big[
  {\bf D}_{\k_\| l;\k_\|+\q'_\|,\lambda}(q'_z)\,
  \rho^{(n-1)}_{\mathrm{irr};\k_\|+\q'_\|,\lambda;\k_\| l}(t_1)
  } \nonumber \\
& & \quad{}- \rho^{(n-1)}_{\mathrm{irr};\k_\| l;\k_\|-\q'_\|,\lambda}(t_1)\,
  {\bf D}_{\k_\|-\q'_\|,\lambda;\k_\| l}(q'_z) \Big]
  \cdot {\bf E}(\q',t_1)
  + \int_{-\infty}^t \!\!\! dt_1\: {\sum_{\kap_\|\lambda}}^\prime
    \gamma_{\k_\| l;\kap_\|\lambda}\,
    \rho^{(n)}_{\mathrm{irr};\kap_\|\lambda;\kap_\|\lambda}(t_1) .
\label{4.rhon.dia1}
\ea
If the contribution of secondary
electrons is small, Eq.~(\ref{4.rhon.odi1}) also applies for $\q_\|=0$,
$l'=l$.

The photoelectron yield for momentum $\k$ and spin $\sigma$ is
given by the time integrated photoelectron current. Equivalently, it can be
written as the occupation of the appropriate vacuum state,
\be
{\cal N}(\k,\sigma) = \rho_{\k\sigma,\ou;\k\sigma,\ou}(t\to\infty) ,
\label{2.N1}
\ee
since electrons are assumed not to leave the states
$|\k\sigma,\ou\rangle$ again.
For the electron states outside of the crystal
\be
\frac{d}{dt}\,\rho_{\k\sigma,\ou;\k\sigma,\ou}
  = \gamma_{\k\sigma,\ou;\k\sigma,\iv} \rho_{\k\sigma,\iv;\k\sigma,\iv} ,
\ee
since their occupation can only change due to
electrons leaving the crystal. With Eq.~(\ref{2.N1}) we find
\be
{\cal N}(\k,\sigma) = \gamma_{\k\sigma,\ou;\k\sigma,\iv}
  \int_{-\infty}^\infty \!\!\! dt\:
  \rho_{\k\sigma,\iv;\k\sigma,\iv}(t) .
\label{2.N2}
\ee
The irreducible contribution to order $n$ is obtained by inserting the
irreducible part of $\rho^{(n)}$, Eq.~(\ref{4.rhon.dia1}),
into this equation. After changing the order of integrals we obtain
\ba
\lefteqn{
{\cal N}^{(n)}_{\mathrm{irr}}(\k,\sigma)
  = \frac{e}{i\hbar}\,
    \frac{\gamma_{\k\sigma,\ou;\k\sigma,\iv}}
    {\Gamma_{\k\sigma,\iv;\k\sigma,\iv}}
  \int_{-\infty}^\infty \!\!\! dt\,
  \sum_{\q\lambda}
  } \nonumber \\
& & {}\times
  \Big[
  {\bf D}_{\k\sigma,\iv;\k_\|+\q_\|,\lambda}(q_z)\,
  \rho^{(n-1)}_{\mathrm{irr};\k_\|+\q_\|,\lambda;\k\sigma,\iv}(t)
  \nonumber \\
& & \;{}- \rho^{(n-1)}_{\mathrm{irr};\k\sigma,\iv;\k_\|-\q_\|,\lambda}(t)\,
    {\bf D}_{\k_\|-\q_\|,\lambda;\k\sigma,\iv}(q_z) \Big]
  \cdot {\bf E}(\q,t) ,
\label{4.Nrho.1}
\ea
where we have used that there is no relaxation {\it into\/} the states above
$\Ev$ in the solid. Note, Eq.~(\ref{4.Nrho.1}) describes photoemission out
of {\it any\/} (possibly non-equilibrium) state.

For ordinary photoemission, ${\cal N}^{(2)}$, we have to calculate the
density operator to first order, $\rho^{(1)}$, which is purely irreducible.
Consequently, the {\it full\/} ordinary photoelectron yield is obtained by
inserting Eq.~(\ref{4.rhon.odi1}) for $n=1$ into Eq.~(\ref{4.Nrho.1}),
\be
{\cal N}^{(2)}(\k,\sigma)
  = \sum_{\q} \int dt_1\,dt_2\,
  \eta_{ij}(\q;t_1,t_2;\k,\sigma)\,E_i(\q,t_1) \,
  E_j(-\q,t_2)
\label{a2.N21}
\ee
with the response function
\ba
\lefteqn{
\eta_{ij}(\q;t_1,t_2;\k,\sigma)
  = \frac{e^2}{\hbar^2} \,
  \frac{\gamma_{\k\sigma,\ou;\k\sigma,\iv}}
    {\Gamma_{\k\sigma,\iv;\k\sigma,\iv}}
  \sum_\lambda
  } \nonumber \\
& & \quad {}\times
  D^i_{\k\sigma,\iv;\k_\|+\q_\|,\lambda}(q_z)\,
  \exp\!\left[i\frac{E_{\k_\|+\q_\|,\lambda}\!-\!E_{\k\sigma,\iv}}{\hbar}
    (t_2-t_1)\right]
  \nonumber \\
& & \quad {}\times
  e^{-\Gamma_{\k_\|+\q_\|,\lambda;\k\sigma,\iv}|t_2-t_1|}\,
  f(E_{\k_\|+\q_\|,\lambda})
  \nonumber \\
& & \quad {}\times
  D^j_{\k_\|+\q_\|,\lambda;\k\sigma,\iv}(-q_z) .
\label{a2.N22}
\ea
It is useful to write our results in the frequency domain. For
ordinary photoemission this yields
\be
{\cal N}^{(2)}(\k,\sigma)
  = \sum_{\q} \int_{-\infty}^\infty\!\!\! d\omega\,
  \eta_{ij}(\q,\omega;\k,\sigma)\,E_i(\q,\omega)\,
  E_j(-\q,-\omega)
\label{4.N2.3}
\ee
with
\ba
\lefteqn{
\eta_{ij}(\q,\omega;\k,\sigma)
  = \frac{2\pi i e^2}{\hbar} \,
  \frac{\gamma_{\k\sigma,\ou;\k\sigma,\iv}}
    {\Gamma_{\k\sigma,\iv;\k\sigma,\iv}}
  } \nonumber \\
& & \; {}\times
  \sum_\lambda
  D^i_{\k\sigma,\iv;\k_\|+\q_\|,\lambda}(q_z)
  \nonumber \\
& & \quad {}\times
  \bigg(\frac1{\hbar\omega+E_{\k_\|+\q_\|,\lambda}-E_{\k\sigma,\iv}
    +i\hbar\Gamma_{\k_\|+\q_\|,\lambda;\k\sigma,\iv}}
  \nonumber \\
& & \qquad
{}- \frac1{\hbar\omega+E_{\k_\|+\q_\|,\lambda}-E_{\k\sigma,\iv}
    -i\hbar\Gamma_{\k_\|+\q_\|,\lambda;\k\sigma,\iv}}\bigg)
  \nonumber \\
& & \quad {}\times
  f(E_{\k_\|+\q_\|,\lambda})\,
  D^j_{\k_\|+\q_\|,\lambda;\k\sigma,\iv}(-q_z) .
\label{4.eta2om.1}
\ea
The third-order contribution, ${\cal N}^{(3)}$, is of interest since the
irreducible third-order response
appears in the reducible contributions to 2PPE.
To calculate ${\cal N}^{(3)}_{\mathrm{irr}}$ from Eq.~(\ref{4.Nrho.1}),
we need the off-diagonal elements of $\rho^{(2)}_{\mathrm{irr}}$ only,
{\it i.e.}, the polarization of the electron system,
which can be obtained from Eq.~(\ref{4.rhon.odi1}) alone.
The result is
\ba
\lefteqn{
{\cal N}^{(3)}_{\mathrm{irr}}(\k,\sigma)
  = \sum_{\q\q'} \int dt_1\,dt_2\,dt_3\,
  \eta^{(3)}_{ijk}(\q,\q';t_1,t_2,t_3;\k,\sigma)
  } \nonumber \\
& & \quad {}\times
  E_i(\q,t_1)\,E_j(\q',t_2)\,E_k(-\q-\q',t_3)
  \qquad\quad
\label{4.N3.3}
\ea
with
\ba
\lefteqn{
\eta^{(3)}_{ijk}(\q,\q';t_1,t_2,t_3;\k,\sigma)
  = \frac{ie^3}{\hbar^3}\,
    \frac{\gamma_{\k\sigma,\ou;\k\sigma,\iv}}
    {\Gamma_{\k\sigma,\iv;\k\sigma,\iv}}
  \sum_{\lambda\lambda'}
  D^i_{\k\sigma,\iv;\k_\|+\q_\|,\lambda}(q_z)\,
  D^j_{\k_\|+\q_\|,\lambda;\k_\|+\q_\|+\q'_\|,\lambda'}(q'_z)\,
  D^k_{\k_\|+\q_\|+\q'_\|,\lambda';\k\sigma,\iv}(-q_z-q'_z)
  } \nonumber \\
& & \quad\times
  \Big(
  \Theta(t_1-t_2)\,\Theta(t_2-t_3)\,
  e^{-i\Omega_{\k_\|+\q_\|,\lambda;\k\sigma,\iv}(t_1-t_2)}
  e^{-i\Omega_{\k_\|+\q_\|+\q'_\|,\lambda';\k\sigma,\iv}(t_2-t_3)}
  \left[-f(E_{\k_\|+\q_\|+\q'_\|,\lambda'})\right]
  \nonumber \\
& & \qquad
{}- \Theta(t_1-t_3)\,\Theta(t_3-t_2)\,
  e^{-i\Omega_{\k_\|+\q_\|,\lambda;\k\sigma,\iv}(t_1-t_3)}
  e^{-i\Omega_{\k_\|+\q_\|,\lambda;\k_\|+\q_\|+\q'_\|,\lambda'}(t_3-t_2)}
  \left[f(E_{\k_\|+\q_\|+\q'_\|,\lambda'})-f(E_{\k_\|+\q_\|,\lambda})\right]
  \nonumber \\
& & \qquad
{}- \Theta(t_3-t_1)\,\Theta(t_1-t_2)\,
  e^{-i\Omega_{\k\sigma,\iv;\k_\|+\q_\|+\q'_\|,\lambda'}(t_3-t_1)}
  e^{-i\Omega_{\k_\|+\q_\|,\lambda;\k_\|+\q_\|+\q'_\|,\lambda'}(t_1-t_2)}
  \left[f(E_{\k_\|+\q_\|+\q'_\|,\lambda'})-f(E_{\k_\|+\q_\|,\lambda})\right]
  \nonumber \\
& & \qquad
{}+ \Theta(t_3-t_2)\,\Theta(t_2-t_1)\,
  e^{-i\Omega_{\k\sigma,\iv;\k_\|+\q_\|+\q'_\|,\lambda'}(t_3-t_2)}
  e^{-i\Omega_{\k\sigma,\iv;\k_\|+\q_\|,\lambda}(t_2-t_1)}
  f(E_{\k_\|+\q_\|,\lambda})
  \Big) .
  \qquad\qquad\qquad\qquad\quad
\label{4.eta3.1}
\ea
Here, $\Omega_{\k l;\k' l'} \equiv (E_{\k l}-E_{\k' l'})/\hbar
- i\Gamma_{\k l;\k' l'}$ is a complex transition frequency.
The four terms in Eq.~(\ref{4.eta3.1}) correspond to different time orders
of interactions with the electric field.
In itself, ${\cal N}^{(3)}$ is usually negligible compared to
${\cal N}^{(2)}$ for the following reason: The three frequencies of
incoming photons have to add up to zero so that one has to
be the negative of the sum of the other two. However, then the sum
frequency is already present in the exciting laser pulse and {\it ordinary\/}
photoemission dominates the signal.

Finally, the irreducible contribution to fourth order has the general form
\ba
\lefteqn{
{\cal N}_{\mathrm{irr}}^{\mathrm{2PPE}}(\k,\sigma)
  = \sum_{\q\q'\q''} \int dt_1\,dt_2\,dt_3\,dt_4\:
  } \nonumber \\
& & \quad {}\times
  \eta^{\mathrm{2PPE}}_{ijkl}(\q,\q',\q'';t_1,t_2,t_3,t_4;
    \k,\sigma)\,
  E_i(\q,t_1)\,E_j(\q',t_2)
  \nonumber \\
& & \quad {}\times
  E_k(\q'',t_3)\,E_l(-\q-\q'-\q'',t_4) .
\label{4.N4.3}
\ea
$\eta^{\mathrm{2PPE}}$ can be found by inserting
Eqs.~(\ref{4.rhon.odi1}) and (\ref{4.rhon.dia1}) into Eq.~(\ref{4.Nrho.1}).
The photoelectron yield is determined by
the off-diagonal components of $\rho^{(3)}$, which in turn depend on
\emph{all} components of $\rho^{(2)}$, including the diagonal ones.
New physics enters here: The 2PPE
current depends on both the polarization and the change of occupation
to second order.

If the increase of the occupation due to secondary electrons is small
to second order, we can use Eq.~(\ref{4.rhon.odi1}) to calculate all
components of $\rho^{(2)}$. The change of occupation due to dipole
transitions and to relaxation {\it out of\/} excited states is included
in Eq.~(\ref{4.rhon.odi1}). Then the response function $\eta^{\mathrm{2PPE}}$ reads
\ba
\lefteqn{
\eta^{\mathrm{2PPE}}_{ijkl}(\q,\q',\q'';t_1,t_2,t_3,t_4;\k,\sigma)
  = \frac{e^4}{\hbar^4}\,
    \frac{\gamma_{\k\sigma,\ou;\k\sigma,\iv}}
    {\Gamma_{\k\sigma,\iv;\k\sigma,\iv}}
  \sum_{\lambda\lambda'\lambda''}
  D^i_{\k\sigma,\iv;\k_\|+\q_\|,\lambda}(q_z)\,
  D^j_{\k_\|+\q_\|,\lambda;\k_\|+\q_\|+\q'_\|,\lambda'}(q'_z)\,
  } \nonumber \\
& & \quad {}\times
  D^k_{\k_\|+\q_\|+\q'_\|,\lambda';
    \k_\|+\q_\|+\q'_\|+\q''_\|,\lambda''}(q''_z)\,
  D^l_{\k_\|+\q_\|+\q'_\|+\q''_\|,\lambda'';\k\sigma,\iv}(-q_z-q'_z-q''_z)
  \nonumber \\
& & \quad {}\times
  \Big[
  F(t_1-t_2,t_2-t_3,t_3-t_4;1,0;2,0;3,0)
  - F(t_1-t_2,t_2-t_4,t_4-t_3;1,0;2,0;2,3) \nonumber \\
& & \qquad{}- F(t_1-t_4,t_4-t_2,t_2-t_3;1,0;1,3;2,3)
  + F(t_1-t_4,t_4-t_3,t_3-t_2;1,0;1,3;1,2) \nonumber \\
& & \qquad{}- F(t_4-t_1,t_1-t_2,t_2-t_3;0,3;1,3;2,3)
  + F(t_4-t_1,t_1-t_3,t_3-t_2;0,3;1,3;1,2) \nonumber \\
& & \qquad{}+ F(t_4-t_3,t_3-t_1,t_1-t_2;0,3;0,2;1,2)
  - F(t_4-t_3,t_3-t_2,t_2-t_1;0,3;0,2;0,1) \Big]
\label{a2.eta41}
\ea
with the auxilliary function
\ba
\lefteqn{
F(\Delta t_1,\Delta t_2,\Delta t_3;
  n_1,n_2;n_3,n_4;n_5,n_6)
  \equiv \Theta(\Delta t_1)\,\Theta(\Delta t_2)\,\Theta(\Delta t_3) }
  \nonumber \\
& & \quad{}\times
  e^{-i\Omega_{n_1,n_2}\Delta t_1}\,
  e^{-i\Omega_{n_3,n_4}\Delta t_2}\,
  e^{-i\Omega_{n_5,n_6}\Delta t_3}\,
  \left[f(E_{n_6})-f(E_{n_5})\right] ,
\label{a2.F1}
\ea
where the states $|n_i\rangle$ are defined as
\ba
|0\rangle \equiv |\k\sigma;\iv\rangle , & \qquad &
  |1\rangle \equiv |\k_\|+\q_\|,\lambda\rangle , \nonumber \\
|2\rangle \equiv |\k_\|+\q_\|+\q'_\|,\lambda'\rangle , & \qquad &
  |3\rangle \equiv |\k_\|+\q_\|+\q'_\|+\q''_\|,\lambda''\rangle .
\ea
Thus the first term in the brackets in Eq.~(\ref{a2.eta41}) reads
\ba
\lefteqn{
\Theta(t-t_1)\,\Theta(t_1-t_2)\,\Theta(t_2-t_3)\,
  e^{-i\Omega_{\k_\|+\q_\|,\lambda;\k\sigma,\iv}(t-t_1)}\,
  e^{-i\Omega_{\k_\|+\q_\|+\q'_\|,\lambda';\k\sigma,\iv}(t_1-t_2)}
  } \nonumber \\
& & \quad{}\times
  e^{-i\Omega_{\k_\|+\q_\|+\q'_\|+\q''_\|,\lambda'';
    \k\sigma,\iv}(t_2-t_3)}
  \big[\underbrace{f(E_{\k\sigma;\iv})}_{=0}
  -f(E_{\k_\|+\q_\|+\q'_\|+\q''_\|,\lambda''})\big]
\ea
etc. $\eta^{\mathrm{2PPE}}$ has the same structure as $\eta^{(3)}$, only with more terms
due to more possible time orders. One should exclude terms from
Eq.~(\ref{a2.eta41}) that correspond
to processes for which the system returns to the equilibrium state after
two of the four interactions. These processes only contribute a small
correction to the numerical prefactor of the ordinary photoelectron yield.

If, in addition, the typical time scale of the experiment, {\it e.g.}, the
delay time in the pump-probe case, is long compared to the dephasing times,
2PPE can be described by the change of occupation alone. Interference
effects are then absent. In this limit, only the terms $F$ in
Eq.~(\ref{a2.eta41}) with $n_3=n_4$ contribute, where $n_3$ is an excited
state reachable by a single interaction out of the Fermi sea. Then
$\eta^{\mathrm{2PPE}}$ simplifies to
\ba
\lefteqn{
\eta^{\mathrm{2PPE}}_{ijkl}(\q,\q',\q'';t_1,t_2,t_3,t_4;\k,\sigma)
  = \frac{e^4}{\hbar^4}\,
    \frac{\gamma_{\k\sigma,\ou;\k\sigma,\iv}}
    {\Gamma_{\k\sigma,\iv;\k\sigma,\iv}}
    \sum_{\lambda\lambda'}
  } \nonumber \\
& & \; {}\times
  \delta_{\q_\|'+\q_\|'',0}\,
  D^i_{\k\sigma,\iv;\k_\|+\q_\|,\lambda}(q_z)\,
  D^j_{\k_\|+\q_\|,\lambda;\k_\|+\q_\|+\q'_\|,\lambda'}(q'_z)
  \nonumber \\
& & \; {}\times
  D^k_{\k_\|+\q_\|+\q'_\|,\lambda';
    \k_\|+\q_\|,\lambda}(-q'_z)\,
  D^l_{\k_\|+\q_\|,\lambda;\k\sigma,\iv}(-q_z)
  \nonumber \\
& & \; {}\times
  \Big[- F(t_1-t_4,t_4-t_2,t_2-t_3;1,0;1,1;2,1) \nonumber \\
& & \quad {}+ F(t_1-t_4,t_4-t_3,t_3-t_2;1,0;1,1;1,2) \nonumber \\
& & \quad {}- F(t_4-t_1,t_1-t_2,t_2-t_3;0,1;1,1;2,1) \nonumber \\
& & \quad {}+ F(t_4-t_1,t_1-t_3,t_3-t_2;0,1;1,1;1,2) \Big] .
\label{ab.e41}
\ea
Note, $\eta^{\mathrm{2PPE}}$ is now proportional to $\exp(-\Gamma_{11}\Delta t)$
from the second exponential in Eq.~(\ref{a2.F1}), where
$\Gamma_{11}=\tau_1^{-1}$ is the {\it energy\/} relaxation rate of state
$|1\rangle=|\k_\|+\q_\|,\lambda\rangle$ and $\Delta t$ is the time between
the second and third interaction. This is easy to understand: After the
second interaction the electron is in the pure state $|1\rangle$,
which decays with the rate $\Gamma_{11}$.

Reducible contributions to the photoelectron current result from
nonlinear optical effects in the solid. They are obtained by replacing
the effective electric field ${\bf E}$ in Eqs.~(\ref{4.N2.3})
and (\ref{4.N3.3}) by the electric field to second order,
see Eq.~(\ref{21.En1}). For 2PPE we obtain the contributions
given in Eqs.~(\ref{22.N2red1}) and (\ref{22.N2red2}) and shown
diagrammatically in Fig.~\ref{fig.N4}(b) and (c). There are also contributions
given by $\eta$ times a product of ${\bf E}$ and
the {\it third-order\/} polarization ${\bf P}^{(3)}$. These are
only significant if the incident light contains a frequency component large
enough to allow ordinary photoemission and we neglect them.

Finally, it is also possible to describe (ordinary) photoemission out of a
general non-equilibrium state described by the density matrix
$\rho_{\mathrm{neq}}$. The irreducible contribution is obtained from
Eq.~(\ref{4.Nrho.1}) for the photoelectron yield ${\cal N}^{(n)}$ by
expressing $\rho^{(n-1)}_{\mathrm{irr}}$ in terms of
the lower-order $\rho^{(n-2)}_{\mathrm{irr}}$ with the help of
Eq.~(\ref{4.rhon.odi1}). This is exact, since only off-diagonal components of
$\rho^{(n-1)}_{\mathrm{irr}}$ are needed. Then $\rho^{(n-2)}$
is replaced by $\rho_{\mathrm{neq}}$.
There is also a reducible part: Two photons can be converted into
a single one at the sum frequency, which for non-equilibrium, when
a finite polarization exists, can lead to a change of occupation of states
above $\Ev$ and thus to photoemission.


\begin{thebibliography}{99}
\bibitem{rem.slow}The dynamics on time scales longer than about
$500\:{\rm fs}$ can be understood by assuming that the electrons have
thermalized at an electronic temperature $T_e$. However, we are interested
in faster effects here.
\bibitem{rem.term2}If $|\k_\| l\rangle$ is a state in the Fermi sea then
the first term in Eq.~(\ref{2.chi210}) corresponds to the process shown in
Fig.~\ref{fig1}, whereas in the second one an electron from deep below
the Fermi energy is excited by the second interaction into the hole left by
the first interaction.
\bibitem{rem.linear}Obviously the system could now emit a photon at the
oscillation frequency. This is the linear optical response described by
$\chi_{ij}$.
\bibitem{rem.param}The fractional numerical parameter values result from
restoration of units and are not in any way special.
\bibitem{rem.SHGband}For a significantly broader intermediate band the
interference falls off slightly faster because processes with larger
detuning obtain more weight relative to those with small detuning.
\bibitem{rem.clusters}Although we treat a flat surface, the general
physical picture applies to clusters as well. To treat clusters
quantitatively in the present framework, one would have to insert the
proper energy eigenstates in the response expressions.
\end{thebibliography}
\end{document}